\documentclass[journal]{IEEEtran}

\usepackage[numbers,sort&compress]{natbib}
\usepackage{url}
\usepackage{bm}
\usepackage{amsfonts}
\usepackage{amsmath}
\usepackage{amssymb}
\usepackage{amsthm}
\usepackage{color}
\usepackage{enumerate}

\usepackage{array,color}%使用表格
\usepackage{tabularx}%使用表格
\usepackage{multirow}%使用多栏宏包
\usepackage{booktabs}%定义了三条划线命令：\toprule、\midrule 和 \bottomrule，可分别对表格顶部、中部和底部使用不同粗细的水平线
\usepackage{makecell}  %table

\usepackage{graphicx}  %figure
\usepackage{subfig}
\usepackage{epstopdf}
\usepackage{graphics}
\usepackage{epsfig}
\usepackage{psfrag}
\usepackage{overpic}

\usepackage{multicol}

 \usepackage{stfloats}
 \usepackage{float}   %强制H

 \usepackage{indentfirst} %缩进
\usepackage{gensymb}

\usepackage{color, colortbl}

\begin{document}

\newcommand {\Data} [1]{\mbox{${#1}$}}  % 数据总格式,不分行

\newcommand {\DataN} [2]{\Data{\Power{{#1}}{{{#2}}}}}  %n维
\newcommand {\DataIJ} [3]{\Data{\Power{#1}{{{#2}\!\times{}\!{#3}}}}}  %ij维

\newcommand {\DatassI} [2]{\!\Data{\Index{#1}{\!\Data 1},\!\Index{#1}{\!\Data 2},\!\cdots,\!\Index{#1}{\!{#2}}}}  % 连续量
\newcommand {\DatasI} [2]{\Data{\Index{#1}{\Data 1},\Index{#1}{\Data 2},\cdots,\Index{#1}{#2},\cdots}}   % 连续量x_1,x_2,...,x_z,...
\newcommand {\DatasII} [3]{\Data{\Index{#1}{{\Index{#2}{\Data 1}}},\Index{#1}{{\Index{#2}{\Data 2}}},\cdots,\Index{#1}{{\Index{#2}{#3}}},\cdots}}  %连续量x_y1,x_y2,...,x_yz,...

\newcommand {\DatasNTt}[3]{\Data{\Index{#1}{{#2}{\Data 1}},\Index{#1}{{#2}{\Data 2}},\cdots,\Index{#1}{{#2}{#3}}} } % 连续量
\newcommand {\DatasNTn}[3]{\Data{\Index{#1}{{\Data 1}{#3}},\Index{#1}{{\Data 2}{#3}},\cdots,\Index{#1}{{#2}{#3}}} } % 连续量

\newcommand {\Vector} [1]{\Data {\mathbf {#1}}}% 向量
\newcommand {\Rdata} [1]{\Data {\hat {#1}}}%生数据
\newcommand {\Tdata} [1]{\Data {\tilde {#1}}} % 经处理过数据
\newcommand {\Udata} [1]{\Data {\overline {#1}}} %均值
\newcommand {\Fdata} [1]{\Data {\mathbb {#1}}} % 域
\newcommand {\Prod} [2]{\Data {\prod_{\SI {#1}}^{\SI {#2}}}}  %累乘
\newcommand {\Sum} [2]{\Data {\sum_{\SI {#1}}^{\SI {#2}}}}   %累加
\newcommand {\Belong} [2]{\Data{ {#1} \in{}{#2}}}  %属于

\newcommand {\Abs} [1]{\Data{ \lvert {#1} \rvert}}  %乘法
\newcommand {\Mul} [2]{\Data{ {#1} \times {#2}}}  %乘法
\newcommand {\Muls} [2]{\Data{ {#1} \! \times \!{#2}}}  %短乘法
\newcommand {\Mulsd} [2]{\Data{ {#1} \! \cdot \!{#2}}}  %短乘法
\newcommand {\Div} [2]{\Data{ \frac{#1}{#2}}}  % 除法
\newcommand {\Trend} [2]{\Data{ {#1}\rightarrow{#2}}}  %趋向于
\newcommand {\Sqrt} [1]{\Data {\sqrt {#1}}} % 根式
\newcommand {\Sqrnt} [2]{\Data {\sqrt[2]{#1}}} % 根式n

\newcommand {\Power} [2]{\Data{ {#1}^{\TI {#2}}}}  % 幂
\newcommand {\Index} [2]{\Data{ {#1}_{\TI {#2}}}}  % 下标

\newcommand {\Equ} [2]{\Data{ {#1} = {#2}}}  % 等于
\newcommand {\Equs} [2]{\Data{ {#1}\! =\! {#2}}}  %短等于
\newcommand {\Equss} [3]{\Equs {#1}{\Equs {#2}{#3}}}  %等于等于

\newcommand {\Equu} [2]{\Data{ {#1} \equiv {#2}}}  %恒等于

\newcommand {\LE}[0] {\leqslant}
\newcommand {\GE}[0] {\geqslant}
\newcommand {\NE}[0] {\neq}
\newcommand {\INF}[0] {\infty}
\newcommand {\MIN}[0] {\min}
\newcommand {\MAX}[0] {\max}

\newcommand {\SI}[1] {\small{#1}}
\newcommand {\TI}[1] {\tiny {#1}}
\newcommand {\Text}[1] {\text {#1}}

%R raw data  T tempt data  X Vector X  x Vector x   sx  x
\newcommand {\VtS}[0]{\Index {t}{\Text {s}}}
\newcommand {\Vti}[0]{\Index {t}{i}}
\newcommand {\Vt}[0]{\Data {t}}
\newcommand {\VLES}[1]{\Index {\tau} {\SI{\Index {}{ \Text{#1}}}}}
\newcommand {\VLESmin}[1]{\Index {\tau} {\SI{\Index {}{ \Text{min\_\Text{#1}}}}}}
\newcommand {\VAT}[0]{\Index {\Vector A}{\Text{Time}}}
\newcommand {\VPbus}[1]{\Index {P}{\Text{Node-}{#1}}}
\newcommand {\VPbusmax}[1]{\Index {P}{\Text{max\_Node-}{#1}}}

\newcommand {\EtS}[2]{\Equs {\Index {t}{\Text {s}}}{#1} {#2}}
\newcommand {\Eti}[2]{\Equs {\Index {t}{i}}{#1} {#2}}
\newcommand {\Et}[2]{\Equs {t}{#1} {#2}}
\newcommand {\EMSR}[2]{\Equs {\Index {\tau} {\SI{\Index {}{ \Text{MSR}}}}}{#1} {#2}}
\newcommand {\EMSRmin}[2]{\Equs {\Index {\tau} {\SI{\Index {}{ \Text{MSR}}}}}{#1} {#2}}

\newcommand {\EAT}[2]{\Equs {\Index {\Vector A}{\Text{Time}}}{#1} {#2}}
\newcommand {\EPbus}[3]{\Equs {\Index {P}{\Text{Node-}{#1}}}{#2} \Text{ #3}}
\newcommand {\EPbusmax}[3]{\Equs {\Index {P}{\Text{max\_Node-}{#1}}}{#2} \Text{ #3}}

\newcommand {\Vgam}[1]{\Index {\gamma}{#1}}
\newcommand {\Egam}[2]{\Equs {\Vgam{#1}}{#2}}

\newcommand {\Emu}[2]{\Equs {{\mu}{#1}}{#2}}
\newcommand {\Esigg}[2]{\Equs {{\sigma}^2{#1}}{#2}}

\newcommand {\Vlambda}[1]{\Index {\lambda}{#1}}

\newcommand {\VV}[1]{\Index {\Vector V}{#1}}
\newcommand {\Vv}[1]{\Index {\Vector v}{#1}}
\newcommand {\Vsv}[1]{\Index {v}{#1}}
\newcommand {\VX}[1]{\Index {\Vector X}{#1}}
\newcommand {\VsX}[1]{\Index {X}{\SI{\Index {}{#1}}}}
\newcommand {\Vx}[1]{\Index {\Vector x}{\SI{\Index {}{#1}}}}
\newcommand {\Vsx}[1]{\Index {x}{\SI{\Index {}{#1}}}}
\newcommand {\VZ}[1]{\Index {\Vector Z}{#1}}
\newcommand {\Vz}[1]{\Index {\Vector z}{\SI{\Index {}{#1}}}}
\newcommand {\Vsz}[1]{\Index {z}{\SI{\Index {}{#1}}}}
\newcommand {\VIndex}[2]{\Index {\Vector {#1}}{#2}}
\newcommand {\VY}[1]{\Index {\Vector Y}{#1}}
\newcommand {\Vy}[1]{\Index {\Vector y}{#1}}
\newcommand {\Vsy}[1]{\Index {y}{\SI{\Index {}{#1}}}}

\newcommand {\VRV}[1]{\Index {\Rdata {\Vector V}}{#1}}
\newcommand {\VRsV}[1]{\Index {\Rdata {V}}{#1}}
\newcommand {\VRX}[1]{\Index {\Rdata {\Vector X}}{#1}}
\newcommand {\VRx}[1]{\Index {\Rdata {\Vector x}}{\SI{\Index {}{#1}}}}
\newcommand {\VRsx}[1]{\Index {\Rdata {x}}{\SI{\Index {}{#1}}}}
\newcommand {\VRZ}[1]{\Index {\Rdata {\Vector Z}}{#1}}
\newcommand {\VRz}[1]{\Index {\Rdata {\Vector z}}{\SI{\Index {}{#1}}}}
\newcommand {\VRsz}[1]{\Index {\Rdata {z}}{\SI{\Index {}{#1}}}}
\newcommand {\VTX}[1]{\Index {\Tdata {\Vector X}}{#1}}
\newcommand {\VTx}[1]{\Index {\Tdata {\Vector x}}{\SI{\Index {}{#1}}}}
\newcommand {\VTsx}[1]{\Index {\Tdata {x}}{\SI{\Index {}{#1}}}}
\newcommand {\VTsX}[1]{\Index {\Tdata {X}}{\SI{\Index {}{#1}}}}
\newcommand {\VTZ}[1]{\Index {\Tdata {\Vector Z}}{#1}}
\newcommand {\VTz}[1]{\Index {\Tdata {\Vector z}}{\SI{\Index {}{#1}}}}
\newcommand {\VTsz}[1]{\Index {\Tdata {z}}{\SI{\Index {}{#1}}}}
\newcommand {\VOG}[1]{\Vector{\Omega}{#1}}

\newcommand {\Sigg}[1]{\Data {{\sigma}^2({#1})}}
\newcommand {\Sig}[1]{\Data {{\sigma}({#1})}}

\newcommand {\Mu}[1]{\Data{{\mu} ({#1})}}
\newcommand {\Eig}[1]{\Data {\lambda}({\Vector {#1}}) }
\newcommand {\Her}[1]{\Power {#1}{\!H}}
\newcommand {\Tra}[1]{\Power {#1}{\!T}}

\newcommand {\VF}[3] {\DataIJ {\Fdata {#1}}{#2}{#3}}
\newcommand {\VRr}[2] {\DataN {\Fdata {#1}}{#2}}

\newcommand {\Tcol}[2] {\multicolumn{1}{#1}{#2} }
\newcommand {\Tcols}[3] {\multicolumn{#1}{#2}{#3} }
\newcommand {\Cur}[2] {\mbox {\Data {#1}-\Data {#2}}}

\newcommand {\VDelta}[1] {\Data {\Delta\!{#1}}}

\newcommand {\STE}[1] {\Fdata {E}{\Data{({#1})}}}
\newcommand {\STD}[1] {\Fdata {D}{\Data{({#1})}}}

\newcommand {\TestF}[1] {\Data {\varphi(#1)}}
\newcommand {\ROMAN}[1] {\uppercase\expandafter{\romannumeral#1}}

\def \FuncC #1#2{
\begin{equation}
{#2}
\label {#1}
\end{equation}
}

\def \FuncCC #1#2#3#4#5#6{
\begin{equation}
#2=
\begin{cases}
    #3 & #4 \\
    #5 & #6
\end{cases}
\label{#1}
\end{equation}
}

\def \Figff #1#2#3#4#5#6#7{   %Label Name1 Name2 Path1 Path2 Caption Form
\begin{figure}[#7]
\centering
\subfloat[#2]{
\label{#1a}
\includegraphics[width=0.23\textwidth]{#4}
}
\subfloat[#3]{
\label{#1b}
\includegraphics[width=0.23\textwidth]{#5}
}
\caption{\small #6}
\label{#1}
\end{figure}
}

\def \Figffb #1#2#3#4#5#6#7#8#9{   %Label Name1 Name2 Name3 Path1 Path2 Path3 Caption Form
\begin{figure}[#9]
\centering
\subfloat[#2]{
\label{#1a}
\includegraphics[width=0.23\textwidth]{#5}
}
\subfloat[#3]{
\label{#1b}
\includegraphics[width=0.23\textwidth]{#6}
}

\subfloat[{#4}]{
\label{#1c}
\includegraphics[width=0.48\textwidth]{#7}
}
\caption{\small #8}
\label{#1}
\end{figure}
}

\def \Figffp #1#2#3#4#5#6#7{   %Label Name1 Name2 Path1 Path2 Caption Form
\begin{figure*}[#7]
\centering
\subfloat[#2]{
\label{#1a}
\begin{minipage}[t]{0.24\textwidth}
\centering
\includegraphics[width=1\textwidth]{#4}
\end{minipage}
}
\subfloat[#3]{
\label{#1b}
\begin{minipage}[t]{0.24\textwidth}
\centering
\includegraphics[width=1\textwidth]{#5}
\end{minipage}
}
\caption{\small #6}
\label{#1}
\end{figure*}
}

\def \Figf #1#2#3#4{   %Label Path1 Caption Form
\begin{figure}[#4]
\centering
\includegraphics[width=0.48\textwidth]{#2}

\caption{\small #3}
\label{#1}
\end{figure}
}

\definecolor{Orange}{RGB}{249,106,027}
\definecolor{sOrange}{RGB}{251,166,118}
\definecolor{ssOrange}{RGB}{254,213,190}

\definecolor{Blue}{RGB}{008,161,217}
\definecolor{sBlue}{RGB}{090,206,249}
\definecolor{ssBlue}{RGB}{200,239,253}

\title{Invisible Units Detection and Estimation Based on Random Matrix Theory}

\author{Xing~He, Lei~Chu, Robert~C. Qiu,~\IEEEmembership{Fellow,~IEEE},  Qian~Ai,   Zenan~Ling, Jian~Zhang
%\thanks{This work was partly supported by N.S.F. of  China  No. 61571296, No.  515771151, and N.S.F.  of  US  Grant No. 1247778, No.  1619250.}

% <-this % stops a space
%\thanks{Xing~He, Robert~C. Qiu, Lei~Chu, and Xinyi~Xu are with the Department of Electrical Engineering, Research Center for Big Data Engineering Technology, State Energy Smart Grid R$\&$D Center, Shanghai Jiaotong University, Shanghai 200240, China. (e-mail: {hexing\_hx@126.com)}}% <-this % stops a space
%\thanks{Robert~C.~Qiu is also with the Department of Electrical and Computer Engineering,
%Tennessee Technological University, Cookeville, TN 38505, USA. (e-mail: {rqiu@tntech.edu})}
%%%National Natural Science Foundation of China
}

\maketitle

%A Methodology of Situation Awareness Based on LES Set and its Spatial and Temporal Performance

\begin{abstract}
Invisible units mainly refer to small-scale units that are not monitored by, and thus are not visible to utilities. Integration of these invisible units into power systems does significantly affect the way in which a distribution grid is planned and operated.
This paper, based on random matrix theory (RMT), proposes a statistical, data-driven framework to handle the massive grid data, in contrast to its deterministic, model-based counterpart.
Combining the RMT-based data-mining framework with conventional techniques, some heuristics are derived as the solution to the invisible units detection and estimation task: linear eigenvalue statistic indicators (LESs) are suggested as the main ingredients of the solution; according to the statistical properties of LESs, the hypothesis testing is formulated to conduct change point detection in the high-dimensional space.
The proposed method is promising for anomaly detection and pertinent to current distribution networks---it is capable of detecting invisible power usage and fraudulent behavior while even being able to locate the suspect's location.
Case studies, using both simulated data and actual data, validate the proposed method.
\end{abstract}


\IEEEpeerreviewmaketitle

\section{Introduction}
\label{Introd}
\IEEEPARstart{F}{uture}  grids will be fundamentally different from current ones \cite{he2015arch}. Technology development, environment pressure, and market reform have greatly spurred the deployment of distributed, renewable, and even plug-and-play units, on both the power generation and the power consumption sides.
% Worldwide small-scale roof-top photovoltaics (PVs) installation reached 23 GW at the end of 2013, and the growth is predicted to be 20 GW per year until 2018~\cite{7426401}. %\cite{epia2014global}.%
% The up-take of electric vehicles (EVs) also continues to increase. At least 665,000 electric-driven light-duty vehicles, 46,000 electric buses, and 235 million electric two-wheelers were in  the worldwide market in early 2015~\cite{parag2016electricity}.
These units are mostly invisible to utilities---they are not monitored by, and thus are not visible to grid operators. 1) Accessing distributed units data into utility systems requires an enormous cost of data acquisition, communication, storage, calculation, and security \cite{7466849}. 2) It is difficult to describe these units using a deterministic model; they are small in size but large in amount, and most of them are with high uncertainty or individuality. 3)  Some behavior, such as power theft~\cite{gaur2016determinants}, unauthorized PV installation~\cite{7426401}, and cyberattack~\cite{Deng2016False}, is essentially invisible.

Lack of visibility may cause incorrect planning and operation of power systems, and even worse, damage to system equipment and customer appliances.
In an environment highly penetrated by invisible units, utilities face technical problems related to overvoltage, frequency control, back feeding flow, and other issues such as a rapid decrease in revenue. %Prosumers like EVs also bring many unknowns and risks that need to be identified and managed \cite{parag2016electricity}.
Moreover, the transparency of the grid is the basis for some advanced management such as dynamic scheduling, demand response.

To enhance the visibility of the network, distribution utilities have begun deploying data collectors such as phasor measurement units (PMUs) \cite{von2014micro}. High resolution voltage and current measurements can be used in a plethora of applications for monitoring, diagnostic, and control purposes, such as state estimation, customer profiling, and anomaly detection \cite{ardakanian2016event}.

In the invisible units detection and estimation task, we deal with a large number (called $N$) of nodes simultaneously. Each node collects a massive data samples (called $T$) for a given period of observation. A classical statistic theory treats a fixed $N$ only. This fixed $N$ is called low-dimensional regime. In practice, we are interested in the case that $N$ can vary arbitrarily in size compared with $T.$ This fundamental requirement is the \textbf{primary driving force} for us to choose RMT as the analysis theory. Indeed, the \textbf{joint distribution} of the eigenvalues is analyzed by RMT as the statistic analytics from big data. To our best knowledge, RMT is developed to address this high-dimensional regime since the classical statistic theory applies to the low-dimensional regime only \cite{qiu2015smart}. Our big data analytics exploit the high-dimensional phenomenon that occurs often in a modern grid. Our special view of angle is expected to contribute some novel algorithms to the community. The connection of our task with RMT may has deep impact on our community.

Lots of work has been done to study the impacts and risks of invisible units %\cite{baran2012accommodating, li2015control, samadi2015static};
\cite{samadi2015static}.
Little attention, however, has been paid to detection and estimation of said invisible units, especially in the context of a complex distributed grid. Some related works are found in the special issue of ``Big Data Analytics for Grid Modernization'' \cite{bda2016tsg}. Reference \cite{7456317} proposes a change-point detection algorithm, which is relevant to our paper to an extent.
Reference \cite{7426401} takes the uncertainty into account, and estimates the power generation of unauthorized PV installation using data generated from a small set of selected representative sites.
Reference \cite{7452675} proposes an approach for anomaly detection and causal impact analysis using a two-layer dynamic optimal synchrophasor measurement devices selection algorithm.
Our previous work \cite{he2015arch, he2015corr, he2016les}, based on RMT, outlines a data-driven framework of big data analytics for power systems.
Our RMT-based framework, \textbf{via spectrum analysis}, studies the \textbf{statistical information which is unique in high-dimensional space}%, e.g., temporal-spatial correlation among the massive datasets
. Besides, the linear eigenvalue statistic (LES) and its statistical properties and advantages, as well as some RMT-relevant operations, are discussed.

\subsection{Contribution}
This paper aims at a challenging and pertinent task to current distribution networks---detection and estimation of invisible units.
Combining the RMT-based data-mining framework with classical hypothesis testing techniques, some heuristics are derived as the solution to the task.
RMT and the relevant operations, which \textbf{perform well in uncertainty processing in high-dimensional space}, are employed to conduct  feature extraction from the massive temporal-spatial data.
Linear eigenvalue statistic indicators (LESs), which are \textbf{robust against data errors} (e.g., data loss, data out-of-synchronization \cite{he2016les}) and \textbf{insusceptible to random noises} (e.g., white noises \cite{he2015arch}),  are employed as the features.
Based on the statistical properties of LESs, a hypothesis testing is formulated to conduct change point (CP) detection for invisible units modeling.
The technical route for the task is given in Fig. \ref{fig:flowpro} in Section \ref{sec:Flowdiag}. The method is model-free and only relies on easily accessible utility data, yet it is effective and fairly powerful---Fig. \ref{Fig:Estimation} in Section \ref{Sec:CaseEsti} shows that a much more accurate result is obtained using our method.

The remainder of this paper is organized as follows.
Section \ref{sec:ResPro} maps the invisible units detection and estimation task onto a mathematical model.
Section \ref{Sec: MathF} studies the mathematical tools to handle the aforementioned model, and then gives a full picture for the task.
Section \ref{section: SIMcase} and \ref{section: Realcase}, with the simulated cases and real-system cases respectively, validate the proposed method.
Section \ref{section: concl} presents the conclusions of this research.

\section{Problem Formulation}
\label{sec:ResPro}
\subsection{Definition of TLPs and ULPs}
This paper attempts to conduct invisible units detection and estimation in a non-omniscient distribution network. More precisely, we aim to obtain users (loads/generators) behavior and accordingly volume at the node level.

From the aspect of users behavior, the loads/generators are divided into two categories: typical load pattern units (TLPs) and uncertain load pattern units (ULPs).
\begin{enumerate}
\item TLPs operate according to well-defined profiles, which are denoted as vectors $\mathbf{p}^{(\rm \alpha)}_i$. \textbf{TLPs' patterns are known in advance---they can be checked by the users based on the daily routine or learned via cluster algorithms using historic data} \cite{chicco2012overview}. For instance, a street-lamp that is set turning on at 18:00 and turning off at 06:00 is a TLP; the street-lamp pattern is modeled as
\[
{{\mathbf{p}}^{\rm \xi}}\!\left(t \right)\!=\!\left\{ \begin{aligned}
 &\text{1}     &t\in\left[ 00:00,06:00 \right]\cup\left[ 18:00,24:00 \right] \\
 &\text{0}     &t\in \left[ 06:00,18:00 \right] \\
\end{aligned} \right..
\]
If the sampling interval is 6 hours, ${\mathbf{p}}^{(\rm \xi)}=[1,0,0,1].$
\item  ULPs are also expressed in time-series, which are denoted as vectors $\mathbf{p}^{(\rm{\beta})}_j$. Comparing to TLPs', \textbf{ULPs' patterns are not accessible in advance or predictable}. The aforementioned invisible units, such as plug-and-play EVs and climate-susceptible PVs, all belong to ULPs.
\end{enumerate}

\subsection{Classification of ULPs}
ULPs can be further divided into three categories---random behavior, invisible behavior, and fraudulent behavior.
Our previous work \cite{he2015arch, he2016les}, based on RMT, has already studied the random behavior, and verified that \textbf{the (independent) random behavior, e.g., white noises, has little impact on the value of LESs}. This paper focuses on the left two---the invisible behavior and the fraudulent one. The examples of the both will be given and studied in Section \ref{Sec:CaseInv}.

The invisible behavior often causes a chain reaction and has an impact on numerous parameters. For instance, the PVs electricity generations do change the power flow of the grid network. The fraudulent behavior, however, often causes parameter deviation alone. For example, some metering errors or power thefts may merely distort the power consumption value $P$ of some node without affecting other variables.
\textbf{The differences between the two behavior above, from the aspect of mathematic, will be revealed by temporal-spatial correlations (among the multiple variables in time-series), which only exist in the high-dimensional space}.

\subsection{Invisible Units Detection and Estimation Task Model}
\label{sec:IUMode}
Considering the volume, we propose to study a general model for each node:
\begin{equation}
\label{Behaviorab}
\underbrace{{{\mathbf{p}}^{(\Sigma )}}}_{\text{Observed Data}}=\underbrace{{{a}_{1}}\mathbf{p}_{1}^{(\rm \alpha)}+\cdots +{{a}_{n}}\mathbf{p}_{n}^{(\rm \alpha)}}_{\text{Power Usage of TLPs}}+\underbrace{{{b}_{1}}\mathbf{p}_{1}^{(\rm \beta)}+\cdots +{{b}_{m}}\mathbf{p}_{m}^{(\rm \beta)}}_{\text{Power Usage of ULPs}},
\end{equation}
where vectors $\mathbf{p}^{(\rm \alpha)}_i$ stand for the daily patterns of TLPs, with the coefficients (volume) $a_i, i\!=\!1,...,n,$ and accordingly, $\mathbf{p}^{(\rm \beta)}_j$ for ULPs, with the coefficients $ b_j,  \; j\!=\!1,...,m.$ As a result, vector $a_i\mathbf{p}^{(\rm \alpha)}_i$ is the daily power usage for the $i$-th TLP, similarly vector $b_j \mathbf{p}^{(\rm \beta)}_j$ for the $j$-th ULP.

\textbf{For the invisible units detection and estimation task, the existence of ULPs need to be detected, and furthermore, the pattern of invisible units $\mathbf{p}^{(\rm \beta)}_j$,  as well as the coefficients (volume) $a_i, b_j$, need to be estimated}.

If all units' pattern and behavior are known in advance, i.e., no $\mathbf{p}^{(\rm \beta)}_j$ exists, or if ULPs are able to be modeled as routine $\mathbf{p}^{(\rm \alpha)}_{n+j}$ instead of uncertain $\mathbf{p}^{(\rm \beta)}_j$, then Eq.~\eqref{Behaviorab} is turned into
\begin{equation}
\label{Behaviora}
\mathbf{p}^{(\Sigma)}=a_1\mathbf{p}^{(\rm \alpha)}_1+a_2\mathbf{p}^{(\rm \alpha)}_2+\cdots+a_{n+m}\mathbf{p}^{(\rm \alpha)}_{n+m}.
\end{equation}

Eq.~\eqref{Behaviora} can be solved through a systematic procedure.
\begin{equation}
\label{Solve1}
\underset{\mathbf{a}}{\mathop{\arg \min }}\,\left\| \mathbf{Pa}-\mathbf{p}^{(\Sigma)} \right\|
\end{equation}
where $\mathbf{P}\!=\!\left[ \begin{matrix}
   {{\mathbf{p}}^{(\rm \alpha)}_{1}} &  \cdots  & {{\mathbf{p}}^{(\rm \alpha)}_{n+m}}  \\
\end{matrix} \right]$, $\mathbf{a}\!=\!{{\left[ \begin{matrix}
   {{a}_{1}}  & \cdots  & {{a}_{n+m}}  \\
\end{matrix} \right]}^{\text T}}$.

Using the least squares method, the estimated value of  $\mathbf{a}$ is obtained as
\begin{equation}
\label{LSPA}
\mathbf{\hat{a}}={{\left( {{\mathbf{P}}^{\text T}}\mathbf{P} \right)}^{-1}}{{\mathbf{P}}^{\text T}}\mathbf{p}^{(\Sigma)}
\end{equation}

It is worth mentioning that the analysis of reactive power $Q$ may be conducted in a similar way.

In a modern distribution network, ULPs $b_j{\mathbf{p}}^{(\rm \beta)}_j$ are present and their influences need to be considered. The existence of ULPs violates the prerequisites of most algorithms (e.g., least squares method) and may cause significant bias on the estimated values of coefficients $a_i$ starting from Eq.~\eqref{LSPA}.
\textbf{Without a fairly correct ULPs detection, spurious results may be obtained}, just as illustrated in Fig. \ref{Fig:Estimation} in Section \ref{Sec:CaseEsti}.

In most scenarios, \textbf{it is reasonable to model $\mathbf{p}^{(\rm \beta)}_j$ as a step signal}. This is the case when EVs charge or PVs generate during $t_a$ to $t_b$. \textbf{Determining the start point $t_a$ and the end point  $t_b$ of the step signal} is at the heart of $\mathbf{p}^{(\rm \beta)}_j$ modeling.

\section{Mathematical Foundation}
\label{Sec: MathF}

\subsection{Random Matrix Theory}
\subsubsection{Statistics based on Random Matrix Theory}
{\Text{\\}}

Random matrices have been an important issue in multivariate statistical analysis since the landmark work of Wishart on fixed size Gaussian matrices. Asymptotic theory on  the limiting spectrum of large random matrices was initially proposed in several works \cite{wigner1958distribution} by Wigner in the 1950s, motivated by problems in quantum physics. Since then, research on  finite spectral analysis of high dimensional random matrices has come under heated discussion by scholars in numerous disciplines \cite{qiu2015smart}. RMT, as a statistical tool with profound theoretical basis, is adapted to multivariate analysis. It can help model many intractable practical systems, especially those with numerous variables. %This paper is mainly concerned with the high-dimensional statistics of random matrices.

\subsubsection{Laws for Spectral Analysis}
{\Text{\\}}

RMT says that for a  Laguerre unitary ensemble (LUE) matrix $\mathbf{A}\!\in\! \VF CNT \left(\Equs {c}{N/T}\le 1\right)$, its  empirical spectral density (ESD) ${g_{\bf{A}}}\left( x \right)$  follows  Marchenko-Pastur (M-P)  Law \cite{marvcenko1967distribution}:
\begin{equation}
\label{eq:D2}
{g_{\bf{A}}}\left( x \right) = \frac{1}{{2\pi cx}}\sqrt {\left( {x - a} \right)\left( {b - x} \right)} , x \in \left[ { a,b} \right]
%&\frac{1}{{2\pi }}\sqrt {4 - {x^2}} &, x \in \left[ { - 2,2} \right]&\text{  ,GUE}; \\
%&\frac{1}{{2\pi cx}}\sqrt {\left( {x - a} \right)\left( {b - x} \right)} &, x \in \left[ { a,b} \right] &\text{  ,LUE};
%\end{aligned} \right.,
\end{equation}
where $a = {\left( {1 - \sqrt c  } \right)^2}, b = {\left( {1 + \sqrt c  } \right)^2}$.

\subsubsection{Universality Principle of  RMT}
{\Text{\\}}

This universality principle \cite{van2014probability}  enables us to obtain the exact asymptotic distributions of various test statistics \textbf{without restrictive distributional assumptions of matrix entries}. For a real system with massive temporal-spatial data, we cannot expect the matrix entries to follow i.i.d. distribution. \textbf{One can perform various hypothesis testings under the assumption that the matrix entries are not Gaussian distributed but use the same test statistics as in the Gaussian case}. Numerous studies using both  grid network data \cite{he2016les}  and power transmission equipment data \cite{yan2018big}  demonstrate that the M-P Law and Ring Law are universally valid---the asymptotic results are \textbf{remarkably accurate for engineering data with relatively moderate matrix sizes such as tens}. This is the very reason why RMT can handle practical massive systems.

\subsection{Connection between RMT and our Task}
\subsubsection{Connection with Random Matrix Theory}
{\Text{\\}}

The observed data, given in Eq.~\eqref{Behaviorab}, consist of multiple variables in time-series. And each variable, i.e., the data of a single node, is a mixture of multiple signals---noises (ULPs), given signals (TLPs), and unknown signals (ULPs).  Under the view of the whole gird network, there should be some high-dimensional statistic information among these signals, \textbf{said temporal-spatial correlations}.
Indeed, further division can be made; for instance, the noises can be divided into white noises and colored noises, and the given signals can be expressed in terms of diverse models such as those based on ARMA processes \cite{yan2018big} and Ornstein-Uhlenbeck processes \cite{perninge2008load}. The division topic will be discussed elsewhere and this paper focuses on the step signal detection as mentioned in Section~\ref{sec:IUMode}, which is an essential anomaly detection task, by utilizing the temporal-spatial data via signal analysis algorithms designing.

The signal analysis algorithms are often designed based on the characteristics of the raw data. For instance, low rank matrix completion (LRMC) are based on low rank \cite{jain2013low}, and sparse principal component analysis (sparse PCA) based on sparsity \cite{zou2006sparse}. Similarly, based on \textbf{the temporal-spatial correlations}, we propose our RMT-based framework.

The signal analysis aiming at extracting the informative statistics (big data analytics) from the massive data, as mentioned in Section \ref{Introd}, \textbf{is a challenge that does not meet the prerequisites of most conventional tools}.
The task has some connection with the field of communication, in particular, massive MIMO (Multiple-Input Multiple-Output) technology in work \cite{zhang2015MassiveMIMO}. In the neighbor communication field, RMT has already been fully proved as an effective and popular tool through hundreds of thousands of published studies. In the field of power systems, however, there are much fewer RMT-relevant researches.

\subsubsection{Connection with Other Potential Algorithms}
{\Text{\\}}

The supervised learning mode is also adapted to handle high-dimensional data.
In a supervised learning context, ground truth data can be exploited to seek parameters automatically. For instance, the state-of-the-art deep learning algorithm does select some non-handcrafted features (called deep features) from the massive labeled dataset without much prior knowledge, so that it can be generalized to different cases without making significant modifications. The deep learning algorithm holds a competitive advantage over our paradigm.
Our paradigm, however, has an advantage of \textbf{transparency}---the algorithm is deeply rooted in RMT.
Unifying time and space through their ratio $c = T /N$, RMT deals with temporal-spatial data \textbf{mathematically rigorously}.
Linear eigenvalue statistics (LESs), built from data matrices, \textbf{follow Gaussian distributions for very general conditions}, and other statistical variables are studied due to the \textbf{latest breakthroughs in probability} on the central limit theorems (CLTs) of those LESs. The statistical properties of these variables are mostly derivable and provable. In this sense, our work is fundamental in nature. Besides, \textbf{RMT performs well with moderate-size (unlabeled) data}, which is often true in engineering.

\subsection{Linear Eigenvalue Statistics and its Central Limit Theorem}
The LES $\tau$ of an arbitrary matrix \Belong{\mathbf{\Gamma}}{\VF CNN} is defined in \cite{lytova2009clrforles,shcherbina2011central}
 via the continuous test function  \Data{\varphi: \Fdata C \rightarrow \Fdata C,}
\begin{equation}
\label{eq:DDLES}
\tau_\varphi=\Sum{i=1}{N}{\varphi({\lambda_i})}=\text{Tr}\varphi \left( \mathbf{\Gamma} \right),
\end{equation}
where the trace of the function of a random matrix is involved.

%It is very interesting to study the special case for a Gaussian random matrix.

\subsubsection{Law of Large Numbers}
{\Text{\\}}

The Law of Large Numbers tells us that $N^{-1}\tau_\varphi$ converges in probability to the limit
\begin{equation}
\label{eq:LES1}
\lim_{N \to \INF}\Div 1N\Sum{i=1}{N}{\varphi({\lambda_i})}\!=\!\int\varphi(\lambda)\rho(\lambda)\,d\lambda,
\end{equation}
{where $\rho(\lambda)$ is the probability density function (PDF) of $\lambda$.}\normalsize{}
%%%%%%%%%%%%%%%%%%%%%%%%%%%%%%%%%%%%%%%%%%%%%%%%%%%%%%%%%%%
%In particular, for the Gaussian orthogonal ensemble (GOE) \cite{lytova2009clrforles} (see  (\ref{eq:GOE1}), (\ref{eq:GOE2}), and (\ref{eq:GOE3})), $\rho(\lambda)$ conforms with the semicircle law:
%\FuncCC {eq:semicirlce density}
%{\rho_{sc}(\lambda)}
%{\Div{1}{2\pi\omega^2}\sqrt{4\omega^2-\lambda^2}}{\lambda^2<4\omega^2}
%{0}  {\lambda^2{\GE}4\omega^2}
%
%The covariance matrix ensemble is another classical type ; M-P Law, as describe in \textit{Section \ref{section: backgroundMath}}, is adapted to this ensemble.
%The covariance matrix is widely used in engineering due to the rectangular form---we can also study  the RMM \Belong{\VX {}}{\VF CNT} with $N\!\ne{\!T}$. We will further discuss this ensemble below.

\subsubsection{Central Limit Theorem}
\label{hearttheory}
{\Text{\\}}

CLT, as the natural second step, aims to study the fluctuations of  LES.
\normalsize{}

\newtheorem{thm55}{Theorem}[section]
\begin{thm55}[M. Sheherbina, 2009, \cite{shcherbina2011central}]
Consider a random matrix $\mathbf{X}\!\in\! {{\mathbb{R}}^{N\!\times\! T}}$  in  ${N\!\times\! T}$ size, and  $\mathbf{M}$ is the covariance matrix \Equs {{\Vector M}}{\Div 1 N\VX{}\VX{}^{\mathrm H}}.
The CLT for {\Vector M} is given as follows:
 Let the real valued test function $\varphi$ satisfy condition ${{\left\| \varphi  \right\|}_{3/2+\varepsilon }}\!<\!\infty   \left( \varepsilon >0 \right)$. Then $\!\tau_\varphi\!$, as defined in Eq.~\eqref{eq:DDLES}, in the limit $N,T\!\to\!\infty , \Equs {c}{N/T}\le 1$, converges in the distribution to the Gaussian random variable with the mean $\mathbb{E}(\tau_\varphi)$, according to Eq.~\eqref{eq:LES1}, and the variance:
\begin{equation}
\label {eq:CLTforLes}
\begin{aligned}
     \sigma^2(\tau_{\varphi})=    &    \frac{2}{c\pi^2 }\iint\limits_{-\frac{\pi }{2}<{{\theta }_{1}},{{\theta }_{2}}<\frac{\pi }{2}}{{{\psi }^{2}}\left( {{\theta }_{1}},{{\theta }_{2}} \right)}\left( 1-\sin {{\theta }_{1}}\sin {{\theta }_{2}} \right) d{{\theta }_{1}}d{{\theta }_{2}} \\
 &                +\frac{{{\kappa}_{4}}}{{\pi }^{2}}\left( \int_{-\frac{\pi }{2}}^{\frac{\pi }{2}}{\varphi \left( \zeta \left( \theta  \right) \right)\sin \theta  d{{\theta }}} \right)^2, \\
\end{aligned}
\end{equation}
{where $\psi \left( {{\theta }_{1}},{{\theta }_{2}} \right)\!=\!\frac{\left[ \varphi \left( \zeta \left( \theta  \right) \right) \right]\arrowvert_{\theta ={{\theta }_{2}}}^{\theta ={{\theta }_{1}}}}{\left[ \zeta \left( \theta  \right) \right]\arrowvert_{\theta ={{\theta }_{2}}}^{\theta ={{\theta }_{1}}}},$
${\left[ {\zeta \left( \theta  \right)} \right]\arrowvert_{\theta  = {\theta _2}}^{\theta  = {\theta _1}}}\!=\!\zeta \left( {{\theta _1}} \right)\! -\! \zeta \left( {{\theta _2}} \right),$
and $\zeta \left( \theta  \right) \!= \!1\! + \!1/c \!+\! {2}\!/\!{\sqrt c} \sin \theta;$   $\kappa_4\!=\!\mathbb{E}\left( {{X_{ij}}^{4}} \right)\! -\!3$ is the $4$-th cumulant of entries of \VX{}.}

\normalsize{}
\label{th555}
\end{thm55}
\normalsize{}

%Our previous work \cite{he2016SAModel} uses Eq.~\eqref{eq:D3} to conduct big data analytics for power systems. This paper takes a fundamentally different approach from Eq.~\eqref{eq:D3}.
To study the convergence as a function of $N,$ we study LES instead of the probability distribution of eigenvalues in Eq.~\eqref{eq:D2}. For an arbitrary test function with enough smoothness, LES $\tau$ (see it as a random variable $Y$) is a positive scalar random variable defined in Eq.~\eqref{eq:DDLES}.   As $N\!\to\! \infty,$ the asymptotic limit of its expectation,   $\mathbb{E}\left( Y \right),$ is given  in Eq.~\eqref{eq:LES1}, and the asymptotic limit of its variance,  $\sigma^2 \left( Y \right),$ is given in Eq.~\eqref{eq:CLTforLes}. \textbf{These two equations are sufficient to study the scalar random variable $Y.$} This approach \textbf{can be viewed as a dimensionality reduction}---the random data matrix of size $N\!\times\! T$ is reduced to a positive scalar random variable $Y$! This dimension reduction is mathematically rigorous only when $N, T\!\to\!  \infty$ but $ \frac{N}{T}\!\to\! c.$ \textbf{Experiences demonstrate, however, that moderate values of $N$ and $T$ are accurate enough for our practical purposes.}

\subsection{LES-based Hypothesis Testing Designing for CP Detection}
\label{sec:IIIc}
Change-point (CP) detection began with Page's (1954, 1955) classical formulation, which was further developed by Shiryaev (1963) and Lorden (1971) \cite{David2013Change}. CP detection is the following problem: suppose $X_1, X_2, \cdots, X_m$ are independent observations. For $j \!\le\! M,$ they have distribution $F_0$; for $j \!>\! M,$ they have distribution $F_1$. The distributions $F_1$ may be completely specified or may depend on unknown parameters.  % In the case of sequential observations we desire to detect the change-point $M$ as soon as possible after it occurs, while rarely claiming a detection before it occurs.
In the case of a fixed number $m$ of observations, we would like to test the null hypothesis of no change, that $F_0\!=\!F_1$, and  to estimate $M$.
%An essential feature of the problem is that $M$ is undefined when $F_0= F_1$, which could also be specified as $M \ge m$ in the fixed sample case or $M = \infty$ in the sequential case.

This paper formulates the hypothesis testing, for the massive dataset, in terms of the statistical properties of LES. As aforementioned, LES, in the limit $N\!,\!T\!\to\!\infty , \Equs {c}{N/T}\!\le\! 1$, converges in the distribution to a Gaussian random variable $\tau_{\varphi}$ with mean $\mathbb{E}(\tau_{\varphi})$, according to Eq.~\eqref{eq:LES1},  and variance $\sigma^2(\tau_{\varphi})$, according to Eq.~\eqref{eq:CLTforLes}.
Moreover, our previous work shows that LES is \textbf{robust against data errors} (e.g., data loss, data out-of-synchronization~\cite{he2016les}) and \textbf{insusceptible to (independent) random noises} (not limited to white noises~\cite{he2015arch}), which is \textbf{not true to those low dimensional statistics} such as mean and variance of any single variable. All of these statistical properties \textbf{make LES a good feature} for a hypothesis testing designing aiming at anomaly detection task.

\textbf{Referring the Gaussian property and standard scores} \cite{wiki2019zscore}\footnote{Standard scores are also called z-values, z-scores, normal scores, and standardized variables. They are most frequently used to compare an observation to a theoretical deviate, such as a standard normal deviate.},  the detection is modeled as a binary hypothesis testing: the normal hypothesis $ {\cal H}_0 $ (no anomaly present) and the abnormal one $ {\cal H}_1 $, denoted by:
\begin{equation}
\label {eq:hypotest}
\left| \begin{aligned}
  & \mathcal{H}_0:  \left| \frac{\tau_{\varphi}- \mathbb{E}(\tau_{\varphi})}{\sigma(\tau_{\varphi})}\right|<\epsilon, \\
 & \mathcal{H}_1:  \left| \frac{\tau_{\varphi}- \mathbb{E}(\tau_{\varphi})}{\sigma(\tau_{\varphi})}\right|\ge\epsilon, \\
\end{aligned} \right.
\end{equation}
where $\epsilon$ is a threshold value that needs to be preset.

It is worth mentioning that the aforementioned Gaussian property and standard scores do offer a reference for setting the threshold value $\epsilon$;
for instance, at a significance level $0.05$, the $\epsilon$ should be set at $1.96$.
However, when setting the threshold range for possible applications in the real world, we should take account of many other factors, e.g., the performance of the real-time $\tau_{\varphi}\!-\!t$ curve, the experiment with the historical data, the experience of the problem solver, and other statistical hypothesis testings such as Student's $t$-test.

\subsection{Matrices Concatenation Operation}
Numerous \textbf{causing factors} affect the \textbf{system state} in different ways. Sensitivity analysis is a valuable and hot topic. The sampling data are in the form of multiple time-series, and we assume that there are $N$ state variables and $M$ factors. Within a fixed period of interest $t_i~(i\!=\!1,\!\ldots\!,T)$, the sampling data of $N$ state variables consist of matrix $\mathbf B\!\in\!\mathbb C^{N \times T}$ (i.e. \textbf{state matrix}), and the factors consist of vector $\mathbf c_j^{\text T} \in \mathbb C^{1 \times T} ~(j\!=\!1,\!\ldots\!,M)$ (i.e. \textbf{factor vector}).
Two matrices with the same length can be put together and thus a new concatenated matrix is formed. In such a way,  matrix $\mathbf A_j$ is formed by concatenating state matrix $\mathbf B$ with factor vector $\mathbf c_j^{\text T}$.

In order to balance the proportion (to increase statistic correlation),  a factor matrix $\mathbf C_j$ is formed by duplicating each factor vector $\mathbf c_j^{\text T}$ for $K$ times\footnote{Empirically,  $K \approx 0.3 \times N.$ \cite{he2015corr}}, written as

\[{{\mathbf{C}}_{j}}={{\left[ \begin{matrix}
   {{\mathbf{c}}_{j}} & {{\mathbf{c}}_{j}} & \cdots  & {{\mathbf{c}}_{j}}
\end{matrix} \right]}^{\text T}}\Belong{}{\mathbb C^{K \times T}}.\]
Then, white noise is introduced into $\mathbf C_j$ to avoid extremely strong cross-correlations. Thus, factor matrix $\mathbf D_j$ for  factor vector $\mathbf c_j^{\text T}$   is expressed as
\begin{equation}
\label{eq: Vec2Matr}
\mathbf D_j=\mathbf C_j+\eta_j \mathbf R \quad (j=1,2,\ldots,m),
\end{equation}
where $\eta_j$ is related to signal-to-noise ratio (SNR), and  matrix $\mathbf R$ is a standard Gaussian Random Matrix.

%Through trace function $\operatorname{Tr}(\cdot),$ SNR of factor matrix $\mathbf D_j$ is defined as
%\begin{equation}
%{\rho _j} = \frac{{\operatorname{Tr} ({{\mathbf{C}}_j}{\mathbf{C}}_j^{\text H})}}{{\operatorname{Tr} ({\mathbf{R}}{{\mathbf{R}}^{\text H}})\eta _j^2}}\quad (j = 1,2, \ldots ,m).
%\end{equation}

In parallel, we construct concatenated matrix $\mathbf A_j$ with each factor $\mathbf c_j^{\text T}$, expressed as
\begin{eqnarray}
\label{eq: a_bc}
\mathbf A_j=\left[
\begin{array}{c}
\mathbf B \\
\mathbf D_j
\end{array}
\right]\in{\mathbb C^{(N+K) \times T}}  \quad (j=1,2,\ldots,m).
\end{eqnarray}

Relationships between causing factors $\mathbf c_j^{\text T}$ and system state $\mathbf B$ can be revealed by  concatenated matrix $\mathbf A_j$.
This concatenated model is compatible with  different units and different measurements for each variable data which are in the form of rows of ${\mathbf A}_j$, due to the normalization during data preprocessing. It is worth mentioning that some mathematical methods, e.g., interpolation,  may be applied to handle sensor data with different sampling rates.

\subsection{Experiment Design for Power Systems}
For power systems, voltage magnitudes $U$ and power consumptions $P$ are preferred for the following reasons: 1) they are easily accessible  and usually at a high accuracy; 2) they belong to measurement parameters, which are independent from network topology;   and 3) our previous work \cite{he2015arch} proves that, for certain scenarios, different types of streaming data, such as $V$ and $I$, may have similar statistical properties in high dimensional space. Similar to Eq.~\eqref{eq: a_bc}, $\mathbf F_j$ is formed as
\begin{eqnarray}
\label{eq: F_UP}
\mathbf F_{j}=\left[
\begin{array}{c}
\mathbf U \\
\mathbf P^{(\Sigma)}_j
\end{array}
\right]\in{\mathbb C^{(N+K) \times T}}
\quad (j=1,2,\ldots,N).
\end{eqnarray}

The state matrix $\mathbf U\in{\mathbb C^{N \times T}}$ consists of voltage magnitudes $[U]_{j,t} (j\!=\!1,\cdots,N,  t\!=\!1,\cdots,T)$, and the $j$-th factor matrix $\mathbf {P}^{(\Sigma)}_j\!\in\!{\mathbb C^{K \times T}}$ consists of active power consumptions $[{P}^{(\Sigma)}_j]_{k,t} (j\!=\!1,\cdots,N, k\!=\!1,\cdots,K, t\!=\!1,\cdots,T)$ according to Eq.~\eqref{eq: Vec2Matr}.

\subsection{Technical route for Detection and Estimation Task}
\label{sec:Flowdiag}
\begin{figure}[hbtp]
\centering
\includegraphics[width=0.42\textwidth]{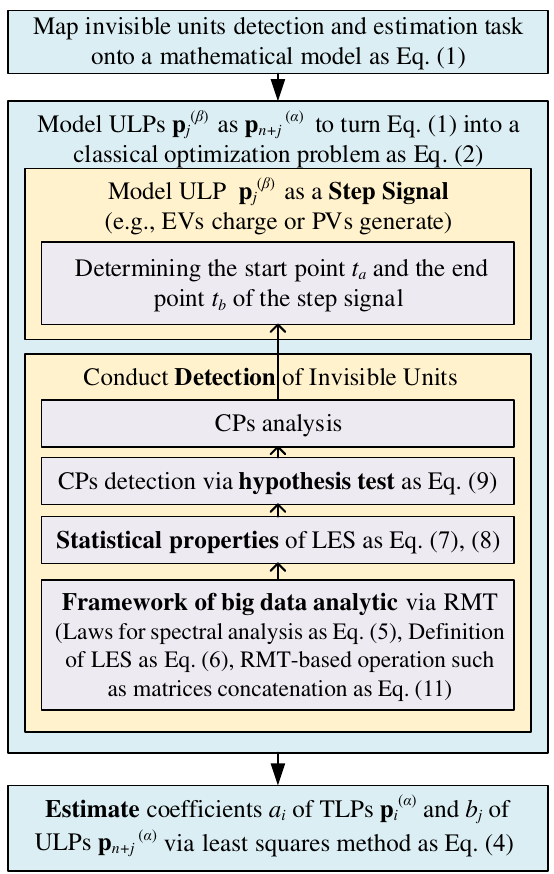}
\caption{Flowchart of RMT-based detection and estimation task}
\label{fig:flowpro}
\end{figure}

Concluding Section \ref{sec:ResPro} and \ref{Sec: MathF}, we outline the flowchart to make a full picture of processes for the invisible units detection and estimation task, as shown in Fig.~\ref{fig:flowpro}.

With the technical route given in Fig.~\ref{fig:flowpro}, we make a progress to our task from the conventional techniques such as change point detection and least squares methods.
\textbf{Employing RMT-based framework, our method conducts big data analytics mathematically rigorously to the massive dataset}.
Based on spectrum analysis, some statistical properties \textbf{which are unique in high-dimensional space} are utilized. For instance, to the statistic LES, its value is robust against data errors and insusceptible to random noises as said in Section \ref{sec:IIIc}, but it is not true to those low dimensional statistics such as mean and variance of any single variable. That is a major reason why we choose RMT-based framework and LESs.

Fig.~\ref{Fig:Estimation} validates the effectiveness of the proposed method---\textbf{a much more accurate estimation of coefficient values for the components is obtained}.

\section{Simulation Cases}
\label{section: SIMcase}
\subsection{Background}
Simulations are based on an IEEE-33 System for a distribution network. For Node $j (j\!=\!1,\cdots,33)$, its gross power usage $\mathbf{p}^{(\Sigma)}_j$ and voltage magnitude $\mathbf{u}_j$ are sampled at a high rate, for example, 9600 points per day (0.11 Hz).
The white noise is added to the power injections as
\begin{equation}\label{eq:gridfluctuation}
{{\tilde y}_{nt}} = {y_{nt}}\left( {1 + {\gamma _1}{z_1}} \right) + {\gamma _2}{z_2},
\end{equation}
where $z_1$ and $z_2$ are two standard Gaussian random variables, i.e. $z_1, z_2\! \sim\! \mathcal{N}\left( {0,1} \right);$  $\gamma_1=0.005, \gamma_2=0.02$.
In this way,
the power flow is obtained via Matpower \cite{MATPOWER2011matpower}.
%\begin{figure}[htbp]
%\centering
%\includegraphics[width=0.48\textwidth, height=0.16\textheight]{Case3IEEE_33.pdf}
%\caption{Topology of the IEEE 33-bus distribution network.}
%\label{fig:case3topo}
%\end{figure}

As mentioned in Sec. \ref{sec:ResPro},  we mainly focus on fraudulent behavior and invisible power usage. Determining the start point and the end point of $\mathbf{p}^{(\rm \beta) }_i$ is the main focus of this study. For longstanding anomalies without sudden change in the observed data segment, some long-term indicators, such as monthly line loss rate, might be effective.
\subsection{Fraud events in a Simple Scenario}
\label{sec:FiSS}
\begin{figure}[hbtp]
\centering
\includegraphics[width=0.49\textwidth]{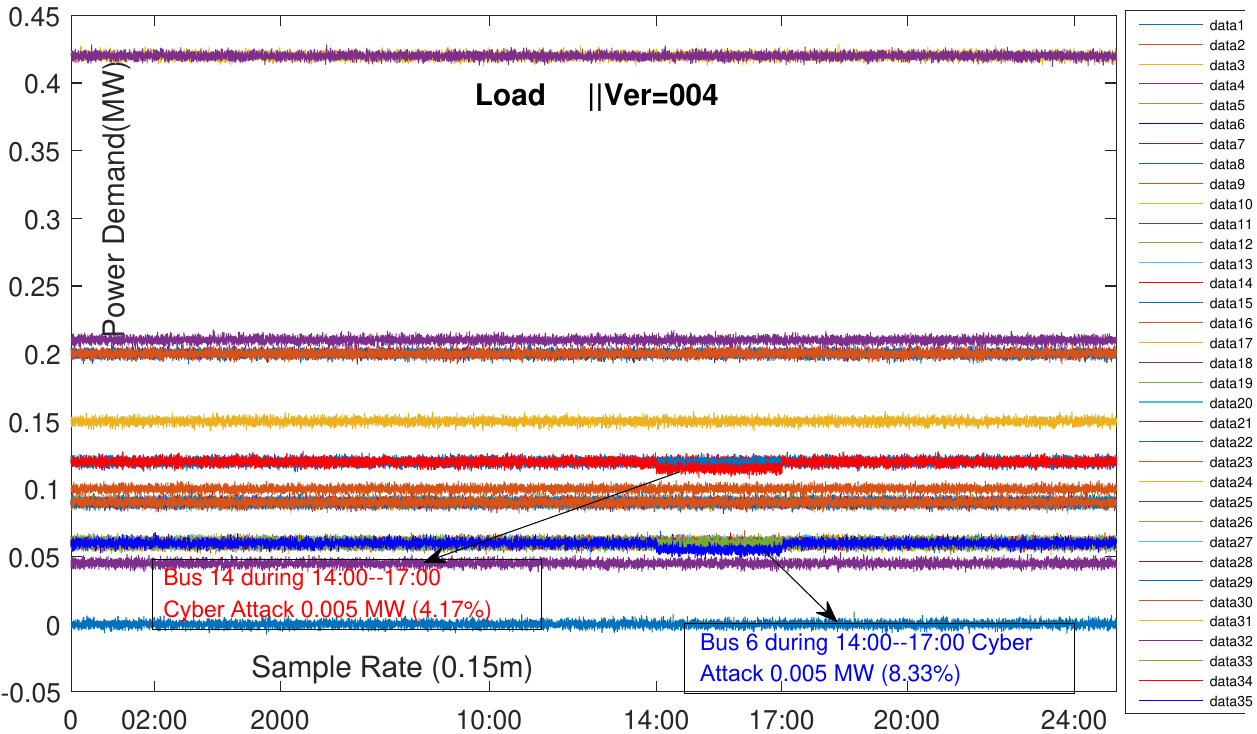}
\caption{Power demand $P$ of each node}
\label{fig:case3A}
\end{figure}

Fraud events often cause parameter deviation alone. Suppose that active power values $P$ for each node stay around their initial points with fluctuations given as Eq.~\eqref{eq:gridfluctuation}. From 14:00 to 17:00, some fraud events on Node 6 and Node 14 cause a reduction of $0.005 \text{ MW}$ ($8.33\%$ of $P_{6}$, $4.17\%$ of $P_{14}$), as shown in Fig. \ref{fig:case3A}.
The lines with legends \textit{data 1} to \textit{data 33} are for the actual power consumption on Node 1 to Node 33, and the lines with legends \textit{data 34} and \textit{data 35} are for the measured power consumption on Node 14 and Node 6, respectively. Note that due to the fraud events, the data with legends \textit{data 14} and \textit{data 6}  are not accessible.

\begin{figure}[hbtp]
\centering
\includegraphics[width=0.49\textwidth]{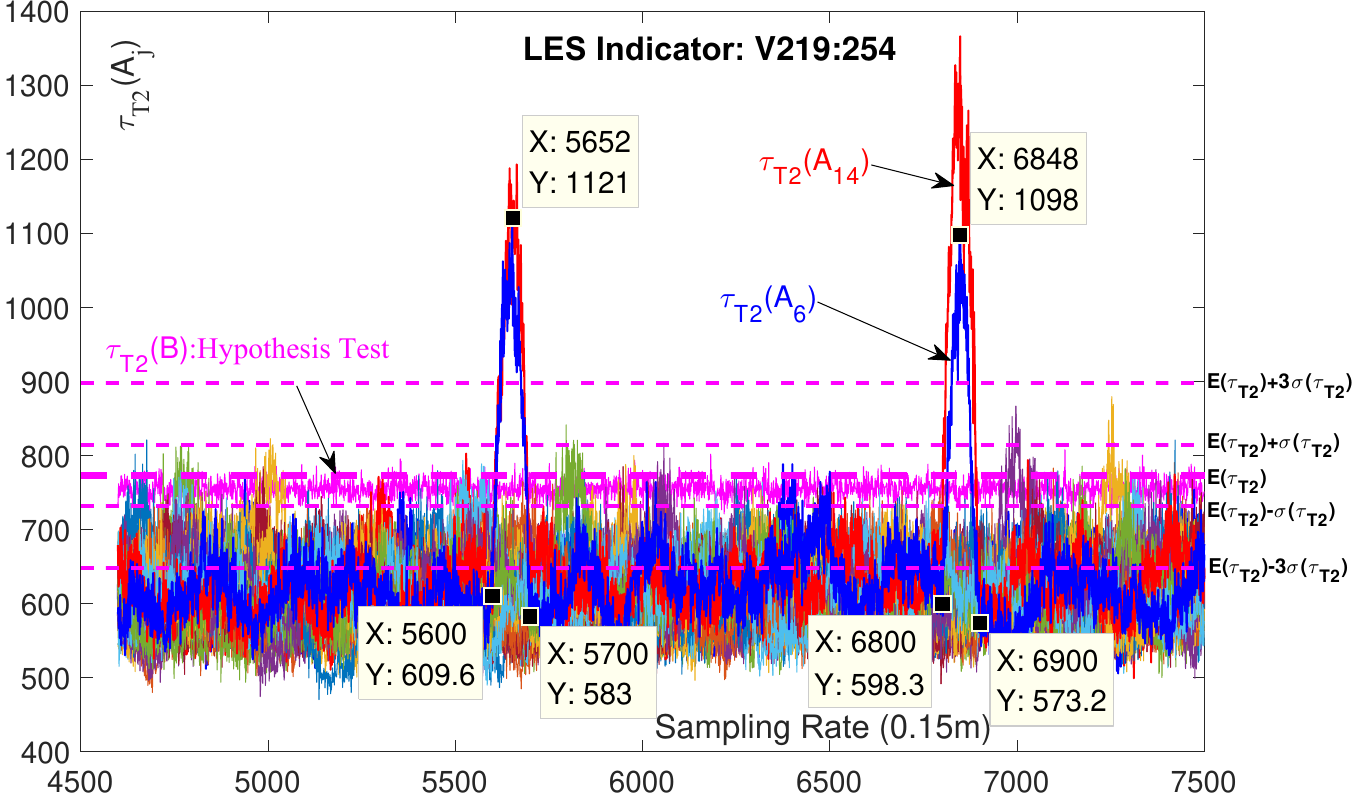}
\caption{LESs in the simple scenario}
\label{fig:case3B}
\end{figure}

Matrices concatenation operation and moving split window analysis are employed for data analysis. According to Eq.~\eqref{eq:DDLES}, we make $N\!=\!33, T\!=\!100, \Delta T\!=\!1,$ and choose Chebyshev polynomials $\text{T}_2$: $ \varphi(x)=2x^2-1$ as the test function.
 LESs $\tau_{\text{T}_2}$ of state matrix $\mathbf B$ and concatenated matrix $\mathbf A_j$ ($j=1,\cdots,33$, referring to Eq.~\eqref{eq: F_UP}) are obtained in Fig.  \ref{fig:case3B}.

In Fig. \ref{fig:case3B},  LES $\tau_{\text{T}_2}$ of state matrix $\mathbf B$, namely, $\tau_{\text{T}_2}(\mathbf B)$ is almost constant. From a statistical view, the theoretical expectation  $\mathbb{E}(\tau_{\text{T}_2})$ and the standard deviation $\sigma(\tau_ {\text{T}_2})$ are accessible via random matrix theory, or rather, via Eq.~\eqref{eq:LES1} and \eqref{eq:CLTforLes}.  It is found that the experimental indicator $\tau_{\text{T}_2}(\mathbf B)$ is exactly bounded between $\mathbb{E}(\tau_{\text{T}_2})\!-\!1.96\sigma(\tau_{\text{T}_2})$ and $\mathbb{E}(\tau_{\text{T}_2})\!+\!1.96\sigma(\tau_{\text{T}_2})$. According to Eq.~\eqref{eq:hypotest}, we cannot reject the null ($\mathcal{H}_0$)---there is no factor actually affecting the system state during the observation period.
On the other hand, the $\tau_{\text{T}_2}$ of state matrix $\mathbf A_i$, namely, $\tau_{\text{T}_2}(\mathbf A_i)$ has four spikes: two spikes for $\tau_{\text{T}_2}(\mathbf A_{6})$ and two spikes for $\tau_{\text{T}_2}(\mathbf A_{14})$. Our previous work \cite{he2015arch} tells us that the anomaly signal should last for $T$ time points (i.e. $T \times 0.15 \text{ minutes}$) and have an  extreme point at $T/2$. This phenomenon is observed  on the $\tau_{\text{T}_2}(\mathbf A_{6})\!-\!t$ curve and $\tau_{\text{T}_2}(\mathbf A_{14})\!-\!t$ curve: $
5700\!-\!5600\!=\!100\!=\!T, \quad 5652\!-\!5600\!\approx50\!=\!T/2.
$

\subsection{Invisible Unit Detection  in a Complex Scenario}
\label{Sec:CaseInv}
%Likelihood Ratio Function  $\text{LR}\!:\!{\varphi_\text{LR}(x)=x-\text{ln}(x)-1}$

This subsection, focusing on fraudulent behavior and invisible behavior, proposes a data-driven solution to the task given in Sec. \ref{sec:ResPro}---\textbf{determining the start point and the end point to model the invisible unit $\mathbf{p}^{(\beta)}_j$ as a step signal.} Firstly, the following assumptions are made:

\begin{enumerate}[(I)]
\item Power usage on Node $j (j\!=\!1,\cdots,33)$ generally consists of four TLPs and one ULP, denoted as
\begin{equation}
\label{eq: PabCase}
\mathbf{p}^{(\Sigma)}_j=a_{j1}\mathbf{p}^{(\rm \alpha)}_{1}+\cdots+a_{j4}\mathbf{p}^{(\rm \alpha)}_{4}+b_{j1}\mathbf{p}^{(\rm \beta)}_1.
\end{equation}
Daily load profiles of TLPs ($\mathbf{p}^{(\rm \alpha)}_{1},\cdots,\mathbf{p}^{(\rm \alpha)}_{4}$) are set as Tab.  \ref{Tab:Case4Load0} and shown in Fig. \ref{fig:Case4Load0}. Note that the blue-filled rectangle indicates that the load profile has a dramatic change at this time point, i.e., change point (CP)  \cite{7456317}.
\item Coefficients $a_i, b_j (i\!=\!1,2,3,4; j\!=\!1)$  are set as Tab~\ref{Tab:Case4Load1}.
\item Invisible power usage events exist on Node 20 and 31: the periods are 1:00--5:00 and 14:00--20:00, and the percentages are 30\% and 50\%, respectively.
\item Fraud events exist on Node 6, 14 and 27: the periods are 20:00--22:00, 14:00--17:00 and 18:00--19:00, and the percentages are 7\%, 8\% and 12\%, respectively.
\end{enumerate}

\begin{figure} [htbp]
\begin{minipage}[b]{1\linewidth} \centering
\setlength{\tabcolsep}{1.7mm}{
\footnotesize
\begin{tabularx}{\textwidth} { >{\scshape}l !{\color{black}\vrule width0.5pt}     >{$}l<{$}  >{$}l<{$}   >{$}l<{$}  >{$}l<{$}  >{$}l<{$}  | >{\scshape}l !{\color{black}\vrule width0.5pt}  >{$}l<{$}  >{$}l<{$}   >{$}l<{$}  >{$}l<{$}  >{$}l<{$}
 }   %OK
\toprule[1.5pt]
\hline
 & \mathbf{p}_1 &   \mathbf{p}_2 & \mathbf{p}_3 & \mathbf{p}_4 & {\mathbf{p}_{\text{u}1}} &      &   \mathbf{p}_1 &   \mathbf{p}_2 & \mathbf{p}_3 & \mathbf{p}_4 & {\mathbf{p}_{\text{u}1}} \\
\hline
\hline
0&	88&	20&	25&	100&	0&      12&	94&	77&	35&	0&	0\\
1&	87&	20&	23&	100&	\multicolumn{1}{>{\columncolor{cyan}}l}{100}&   13&	\multicolumn{1}{>{\columncolor{cyan}}l}{86}&	80&	30&	0&	0\\
2&	88&	20&	22&	100&	100&     14&	86&	86&	33&	0&	\multicolumn{1}{>{\columncolor{cyan}}l}{100}\\
3&	\multicolumn{1}{>{\columncolor{cyan}}l}{100}&	21&	22&	100&	100&     15&	88&	86&	44&	0&	100\\
4&	96&	20&	27&	100&	100&   16&       	85&	87&	50&	\multicolumn{1}{>{\columncolor{cyan}}l}{100}&	100\\
5&	100&	20&	31&	100&	\multicolumn{1}{>{\columncolor{cyan}}l}{0}     &  17&	87&	\multicolumn{1}{>{\columncolor{cyan}}l}{35}&	56&	100&	100\\
6&	98&	20&	29&	\multicolumn{1}{>{\columncolor{cyan}}l}{0}&	0  &  18&	88&	25&	\multicolumn{1}{>{\columncolor{cyan}}l}{85}&	100&	100\\
7&	97&	30&	28&	0&	0&       19&	85&	25&	80&	100&	100\\
8&	\multicolumn{1}{>{\columncolor{cyan}}l}{88}&	\multicolumn{1}{>{\columncolor{cyan}}l}{40}&	31&	0&	0&  20&	84&	20&	70&	100&	\multicolumn{1}{>{\columncolor{cyan}}l}{0}\\
9&	82&	\multicolumn{1}{>{\columncolor{cyan}}l}{85}&	37&	0&	0&       21&	83&	20&	76&	100&	0\\
10&	82&	85&	42&	0&	0&     22&	86&	20&	\multicolumn{1}{>{\columncolor{cyan}}l}{43}&	100&	0\\
11&	\multicolumn{1}{>{\columncolor{cyan}}l}{95}&	82&	42&	0&	0&      23&	88&	15&	30&	100&	0\\
\hline
\toprule[1pt]
%\end{tabular*}
\end{tabularx}
}
\raggedright
\scriptsize {Note: blue-filled rectangle indicates CP.}
\normalsize{}
\captionof{table}{TLPs, ULPs and their 24-hour power demands.}
\label{Tab:Case4Load0}
\end{minipage}
\end{figure}

\begin{table}[hbp]
\caption{Coefficients of TLPs and ULP of each node.}
\label{Tab:Case4Load1}

\begin{minipage}[!h]{0.5\textwidth}
\centering

\footnotesize
%\scriptsize
%\tiny
\setlength{\tabcolsep}{1.3mm}{
\begin{tabularx}
{1\textwidth}
{ >{\scshape}l !{\color{black}\vrule width0.5pt}      >{$}l<{$}  >{$}l<{$}   >{$}l<{$}  >{$}l<{$}  >{$}l<{$}|!{\color{black}\vrule width1pt}  >{\scshape}l !{\color{black}\vrule width0.5pt}  >{$}l<{$}  >{$}l<{$}   >{$}l<{$}  >{$}l<{$}  >{$}l<{$}
 }   %OK
\toprule[1.5pt]
\hline
{\text{}} & a_1 & a_2 & a_3 & a_4 & b_1 &{\text{}} & a_1 & a_2 & a_3 & a_4 & b_1 \\
\hline
\hline
1&	0.25&	0.25&	0.25&	0.25&	0& 2&	0&	0.7&	0.1&	0.2&	0\\
3&	0&	0.1&	0.8&	0.1&	0& 4&	0.05	&0.75	&0.1	&0.1&	0\\
5&	0&	0.1	&0.8	&0.1&	0& 6&	0.1&	0.2	&0.5	&0.2&	0\\
7&	0.8	&0.05&	0.1&	0.05&	0 &8&	0.85&	0.05&	0&	0.1	&0\\
9&	0.1&	0.15	&0.6	&0.15	&0& 10&	0&	0.15	&0.8&	0.05&	0\\
11&	0&	0.2&	0.75&	0.05	&0&12&	0.05&	0.1&	0.75&	0.1&	0\\
13&	0.05&	0.05&	0.85&	0.05&	0&14&	0.7&	0.05&	0.2&	0.05&	0\\
15&	0&	0.05	&0.9	&0.05&	0& 16&	0&	0	&0.95&	0.05&	0\\
17&	0&	0.1	&0.8&	0.1&	0& 18&	0&	0.7&	0.1&	0.2&	0\\
19&	0&	0.5	&0.1&	0.4&	0& 20&	0&	0.2	&0.2&	0.3&	0.3\\
21&	0&	0.8	&0.1&	0.1&	0& 22&	0.1&	0.75	&0&	0.15&	0\\
23&	0.2&	0.6	&0&	0.2&	0& 24&	0.85&	0&	0.05&	0.1&	0\\
25&	0.75&	0.1&	0.1&	0.05&	0 &26&	0.2&	0&	0.7&	0.1&	0\\
27&	0.1&	0	&0.75&	0.15&	0 &28&	0.25&	0.1&	0.6&	0.05&	0\\
29&	0.8&	0.05&	0.1	&0.05&	0& 30&	0.9&	0&	0.05&	0.05&	0\\
31&	0.1&	0.1&	0.05	&0.25&	0.5& 32&	0.9&	0&	0&	0.1	&0\\
33&	0.95&	0&	0&	0.05&	0& & & & & &\\
\hline
\toprule[1pt]
%\end{tabular*}
\end{tabularx}
}
\normalsize{}
\end{minipage}
\end{table}

\begin{figure*}[htbp]
\centering
\subfloat[Active Power $P_{\Sigma  k,j}$  of each node]{\label{fig:case4C10}
\includegraphics[width=0.49\textwidth]{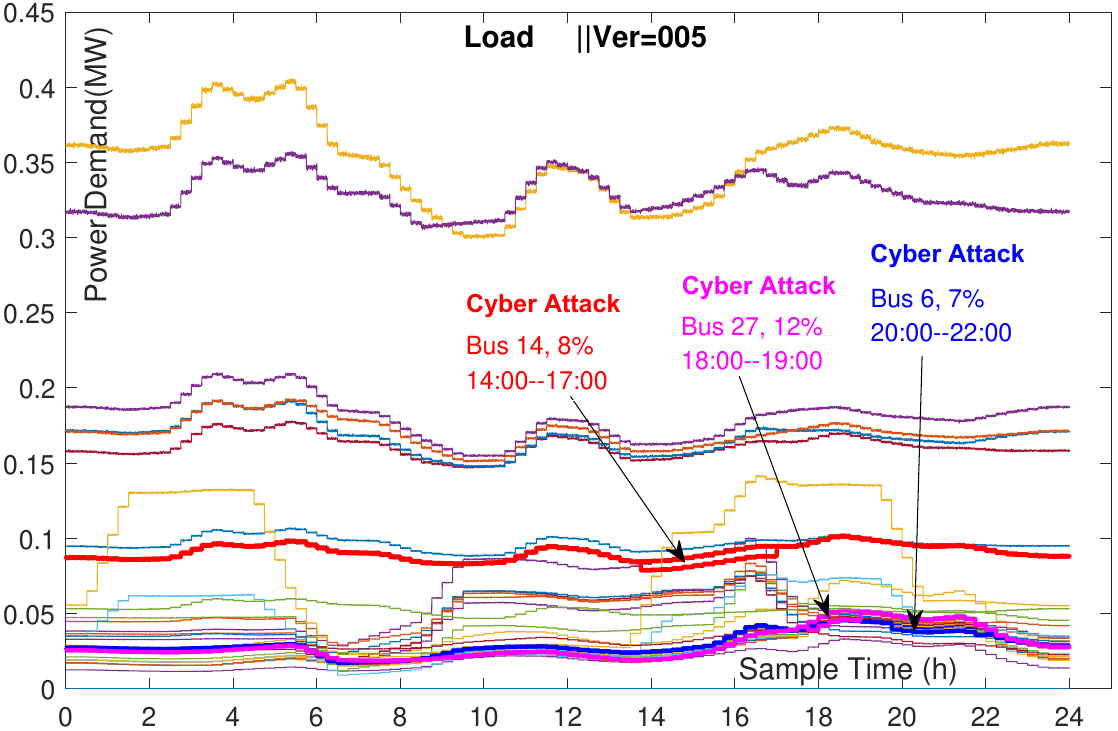}
}
\subfloat[Voltage $U_{i,j}$ of each node]{\label{fig:case4C2}
\includegraphics[width=0.49\textwidth]{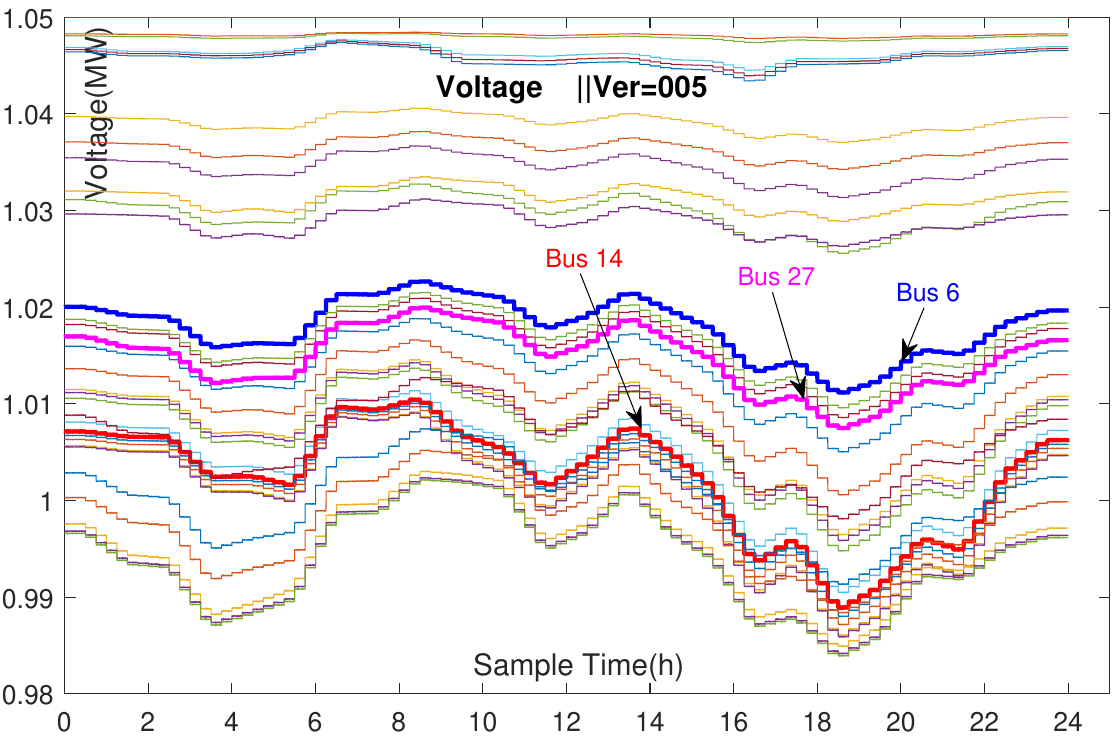}
}

\subfloat[$t-\tau_{\text{T}_2}$ Curve of each node]{\label{fig:case4C11}
\includegraphics[width=0.49\textwidth]{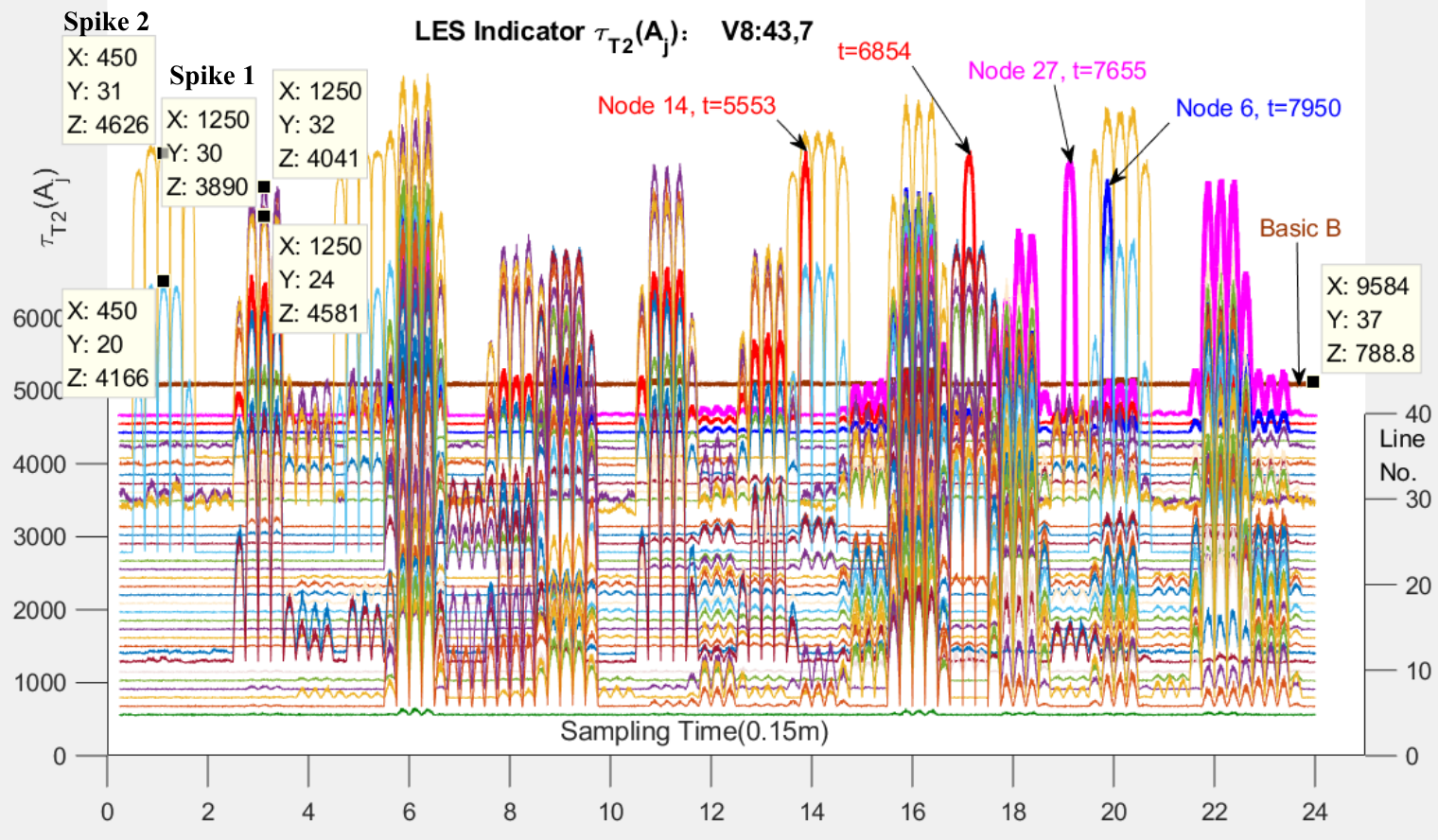}
}
\subfloat[$t-\tau_{\text{T}_2}$ Curve of Node 6, 14, and 27]{\label{fig:case4D}
\includegraphics[width=0.49\textwidth]{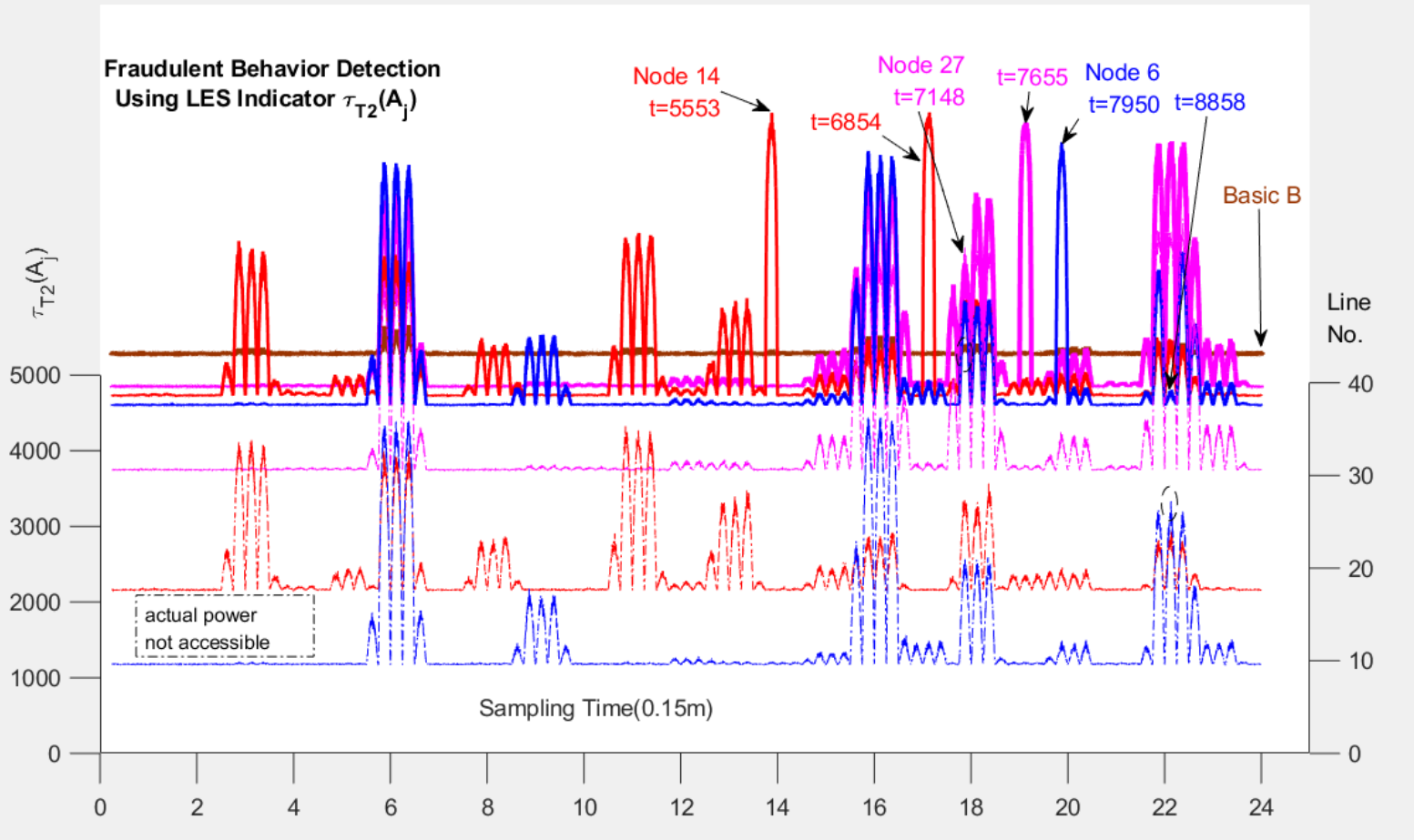}
}
\caption{Illusion of the data and analysis  of a complex scenario for behavior analysis.}
\label{fig:case3C}
\end{figure*}

In the assumed complex scenario above, the active power $P$ and the voltage $U$ of each node are accessible, as shown in Fig. \ref{fig:case4C10} and \ref{fig:case4C2}, respectively. Following Eq.~\eqref{eq: F_UP}, $\mathbf F_j$ is obtained for Node $j (j\!=\!1,\cdots,36)$. %It is worth mentioning that among these 36 vectors, 33 vectors are for measured power consumption data, and 3 vectors are for  extra real power consumption data which are not accessbile.
The LESs of $\mathbf F_j$ are obtained as the $\tau_{\text{T}_2}\!-\!t$ curves shown in Fig. \ref{fig:case4C11} and Fig. \ref{fig:case4D}.

These figures give a visualization of LESs throughout a whole day. \textbf{Particular attention should be paid to the spikes---their corresponding time point and line number hold vital clues about when and where the behavior occurs}:

\begin{itemize}
\item The brown line  indicates $\tau_{\text{T}_2}(\mathbf B)$; it is relatively smooth because the system is stable without emergencies such as voltage collapse. Our previous work \cite{he2015arch} verifies that the independent random noises ,e.g., completely random behavior, has little impact on the value of LESs.
\item For fraud events, the extreme points are located at $t\!=\! 5553, 6854,  7655,$ etc. According to Section \ref{sec:FiSS}, this phenomenon \textbf{matches Assumption \MakeUppercase{\expandafter\romannumeral4}} that the CPs are at $t\!=\! 14\!:\!00 (5600\!\approx\!5553\!+\!50), 17\!:\!00 (6800\!\approx\!6854\!-\!50), 19\!:\!00 (7600\!\approx\!7655\!-\!50)$, etc., respectively.
\item For TLPs, \textit{Spike 1 (X: 1250, Y: 32, Z: 4041)} indicates the existence of a CP at $3:00 (1200\!=\!1250\!-\!50)$ on Node 32. According to Tab. \ref{Tab:Case4Load0}, 3:00 is a CP of TLP $\mathbf{p}^{(\rm \alpha)}_1.$ Thus we can deduce that  TLP $\mathbf{p}^{({\alpha})}_1$  takes a dominant on Node 32. Similar deductions can be made for Node 25, 24, 30, etc. These deductions  \textbf{match Assumption are confirmed by Tab. \ref{Tab:Case4Load1}}.
\item For invisible ULPs, \textit{Spike 2 (X: 450, Y: 31 Z:4626)} indicates the existence of a CP at $1:00 (400\!=\!450\!-\!50)$ on Node 31. According to Tab. \ref{Tab:Case4Load0},  there is no existence of any TLP ($\mathbf{p}^{(\rm \alpha)}_1$, $\mathbf{p}^{(\rm \alpha)}_2$, $\mathbf{p}^{({\rm \alpha})}_3$, $\mathbf{p}^{({\rm \alpha})}_4$) matching CP 1:00. As a result, we artificially build a CP at 1:00 for a new ULP ($\mathbf{p}^{(\rm \beta)}_1$ for this case). Step by step, all the CPs of ULP $\mathbf{p}^{(\rm \beta)}_{1}$ are obtained as Tab. \ref{Tab:Case4Load0}. Then, the general model Eq.~\eqref{eq: PabCase}  is tuned into the classical model Eq.~\eqref{Behaviora} $\mathbf{p}^{(\Sigma)}=a_1\mathbf{p}^{(\rm \alpha)}_1+a_2\mathbf{p}^{(\rm \alpha)}_2+a_3\mathbf{p}^{(\rm \alpha)}_3+a_4\mathbf{p}^{(\rm \alpha)}_4+a_5\mathbf{p}^{(\rm \alpha)}_5$.
\end{itemize}

\subsection{Estimation with and without Invisible Units Detection}
\label{Sec:CaseEsti}
\begin{figure}[htpb]
\centering
\includegraphics[width=0.46\textwidth,height=0.22\textheight]{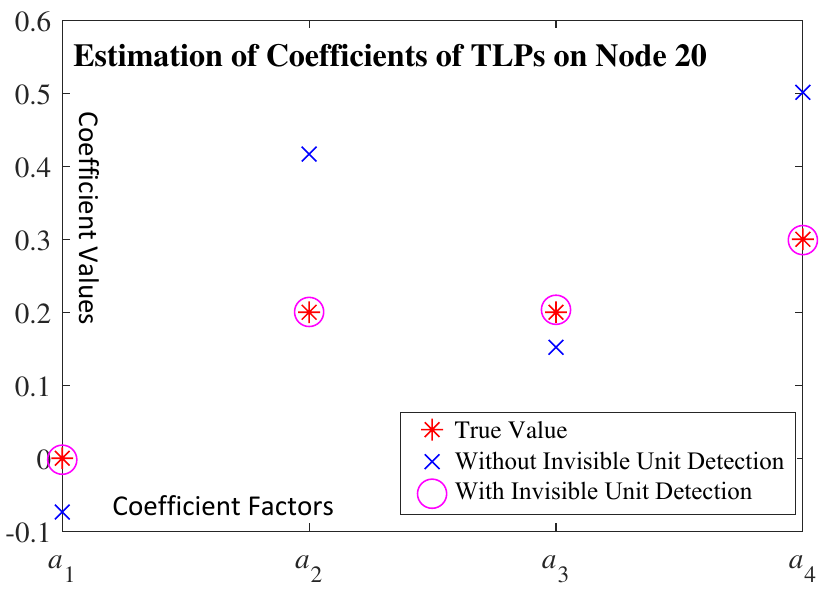}
\caption{Estimate with and without Invisible Units Detection}
\label{Fig:Estimation}
\end{figure}

Existence of invisible units disables the classical least squares method. Taking Node 20 for instance, without correct detection of the invisible units, we will get a bad result as shown in  Fig.  \ref{Fig:Estimation}. However, if the information about the start point and end point of the invisible behavior is acquired, an accurate estimate can be obtained.

\section{Real-World Case Studies}
\label{section: Realcase}

\subsubsection{Data}  
\begin{figure*}[!t]
\centering
\begin{minipage}{.2\linewidth} \centering
\subfloat[The Grid Network]{\label{Fig:Case5Gird}
\includegraphics[width=1\textwidth,height=0.15\textheight]{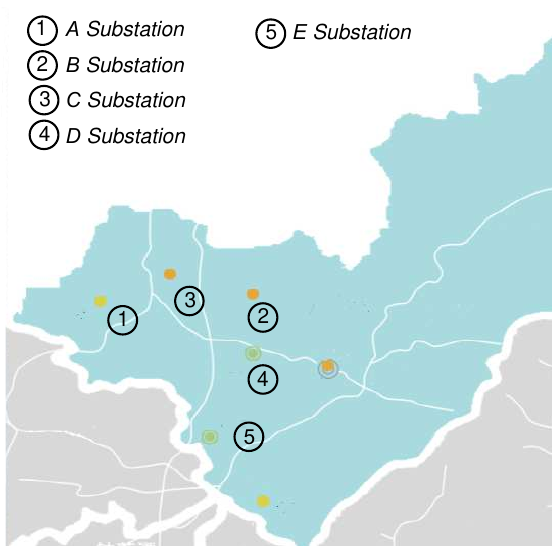}
}

\subfloat[Ring Law of $\mathbf {X}_0$]{\label{fig:case5Ring}
\includegraphics[width=1\textwidth,height=0.15\textheight]{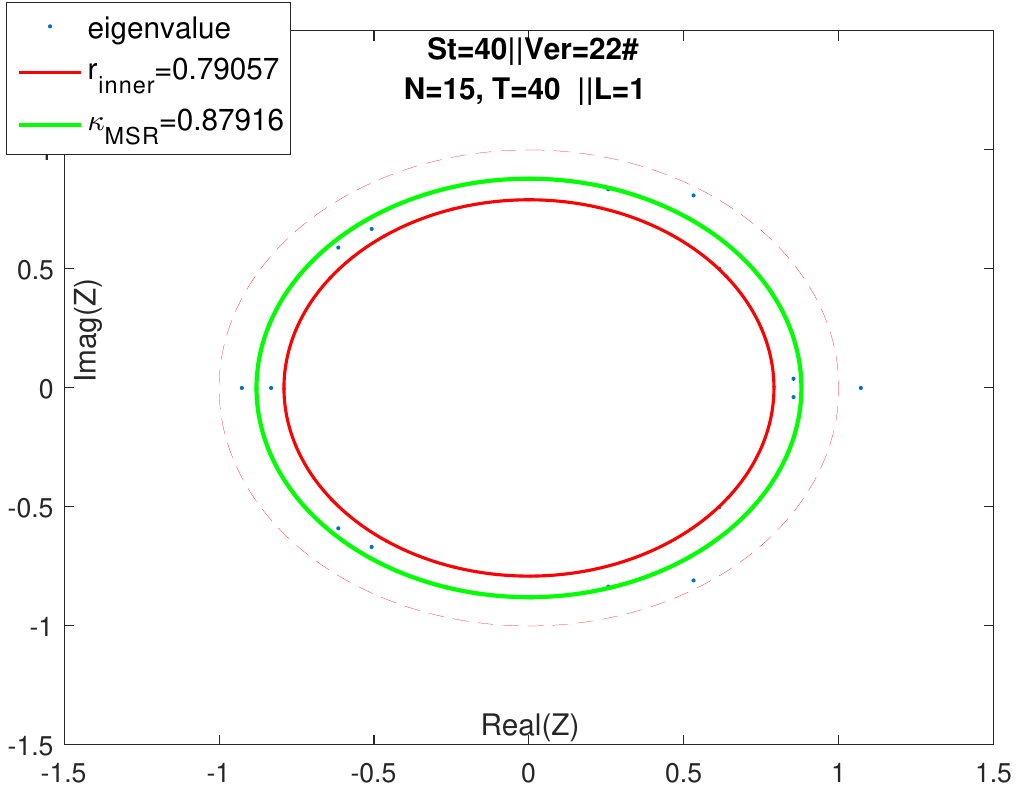}
}
\end{minipage}
\begin{minipage}{.36\linewidth} \centering
\subfloat[Day 1: Voltage $V$]{\label{fig:case5A1}
\includegraphics[width=1\textwidth,height=0.15\textheight]{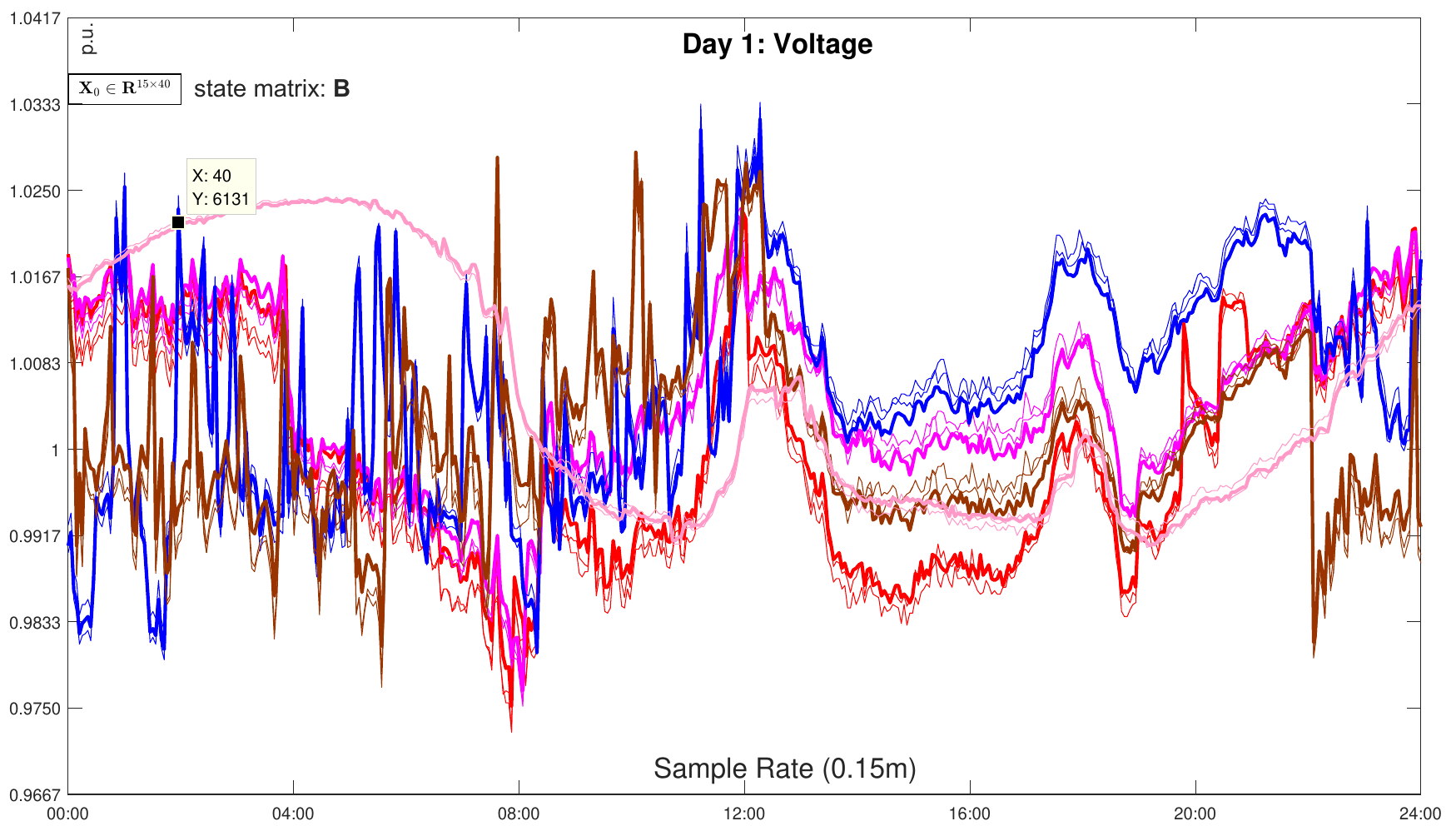}
}

\subfloat[Day 2: Voltage $V$]{\label{fig:case5A2}
\includegraphics[width=1\textwidth,height=0.15\textheight]{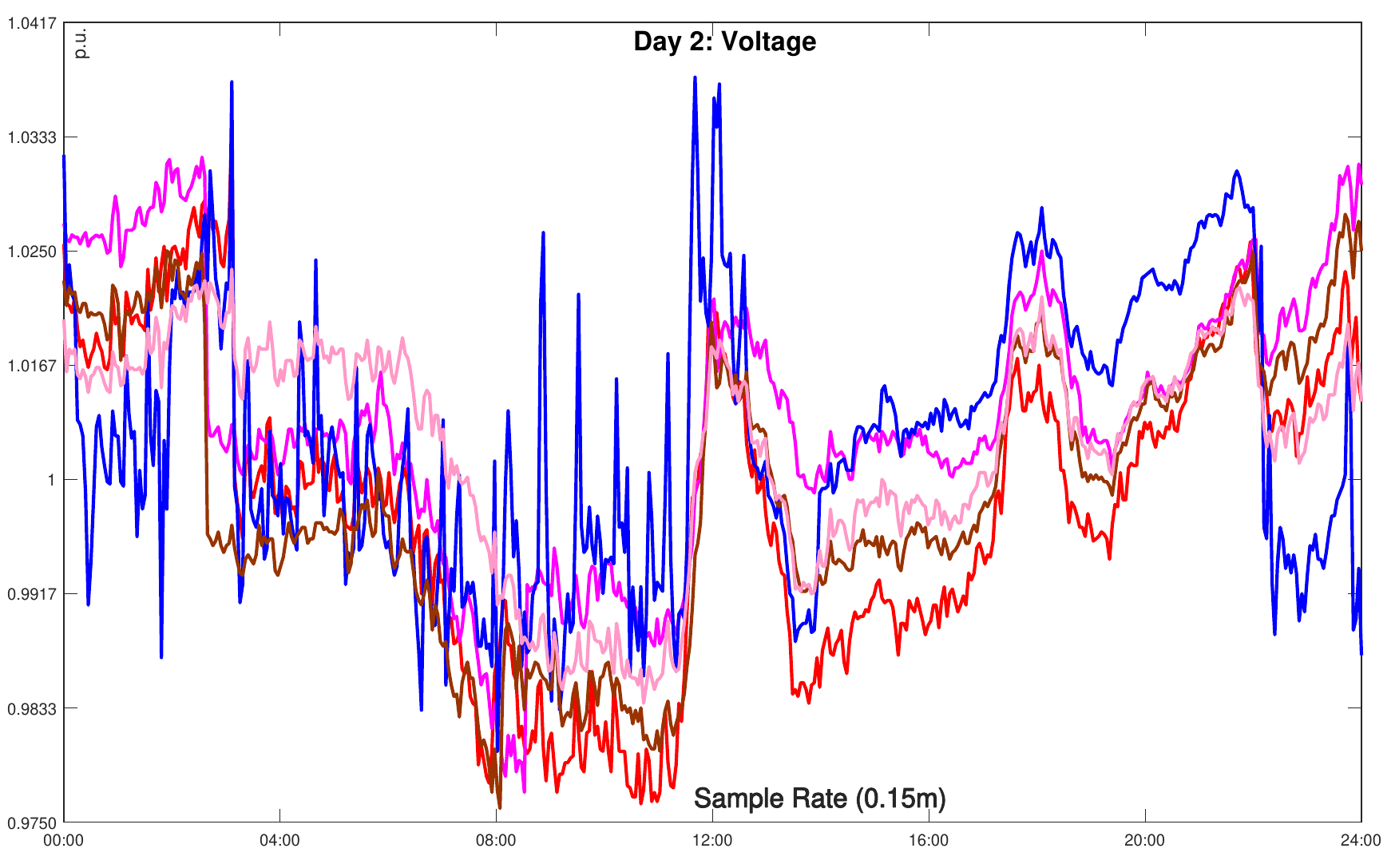}
}
\end{minipage}
\begin{minipage}{.36\linewidth} \centering
\subfloat[Day 1: Current $I$]{\label{fig:case5B1}
\includegraphics[width=1\textwidth,height=0.15\textheight]{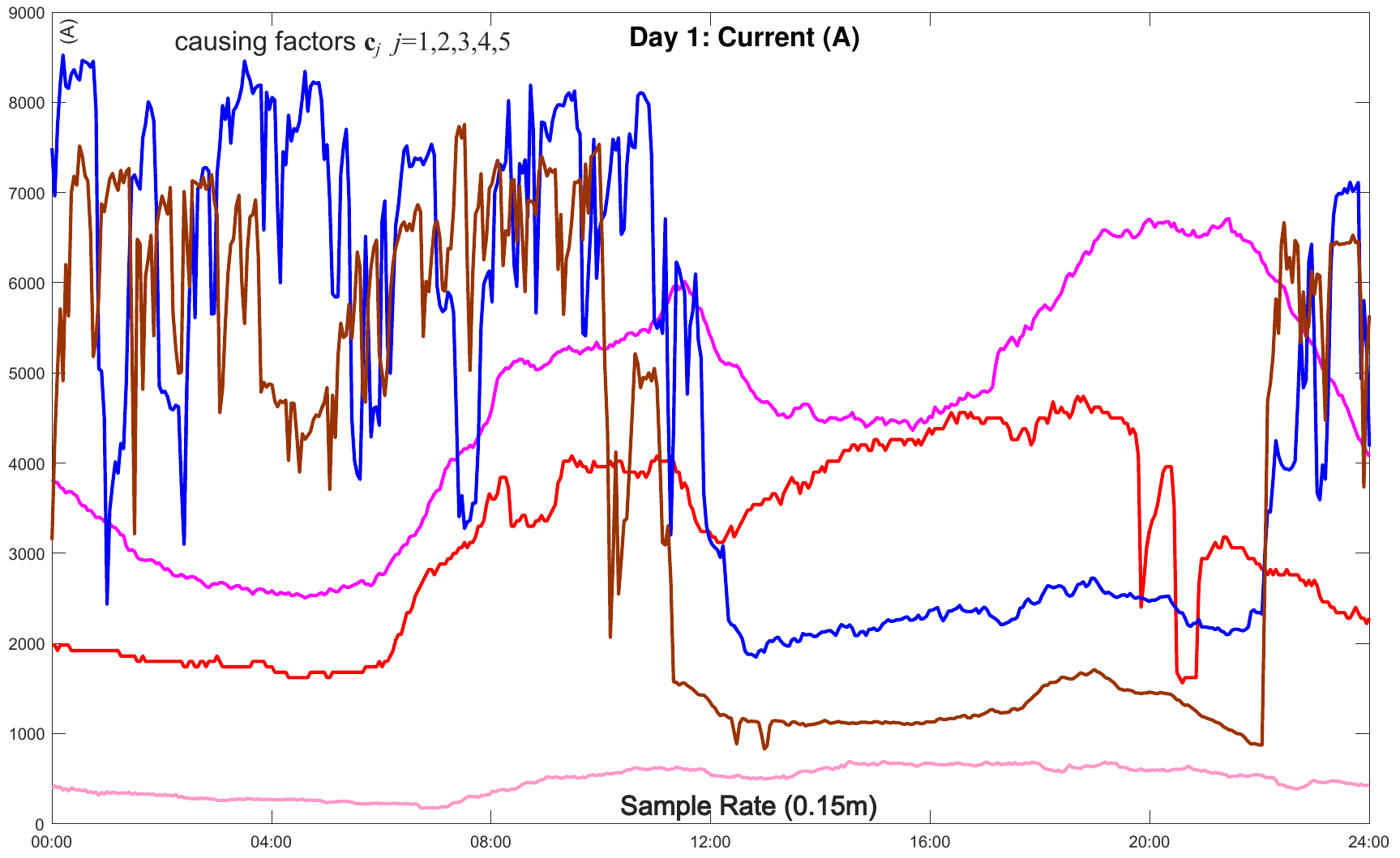}
}

\subfloat[Day 2: Current $I$]{\label{fig:case5B2}
\includegraphics[width=1\textwidth,height=0.15\textheight]{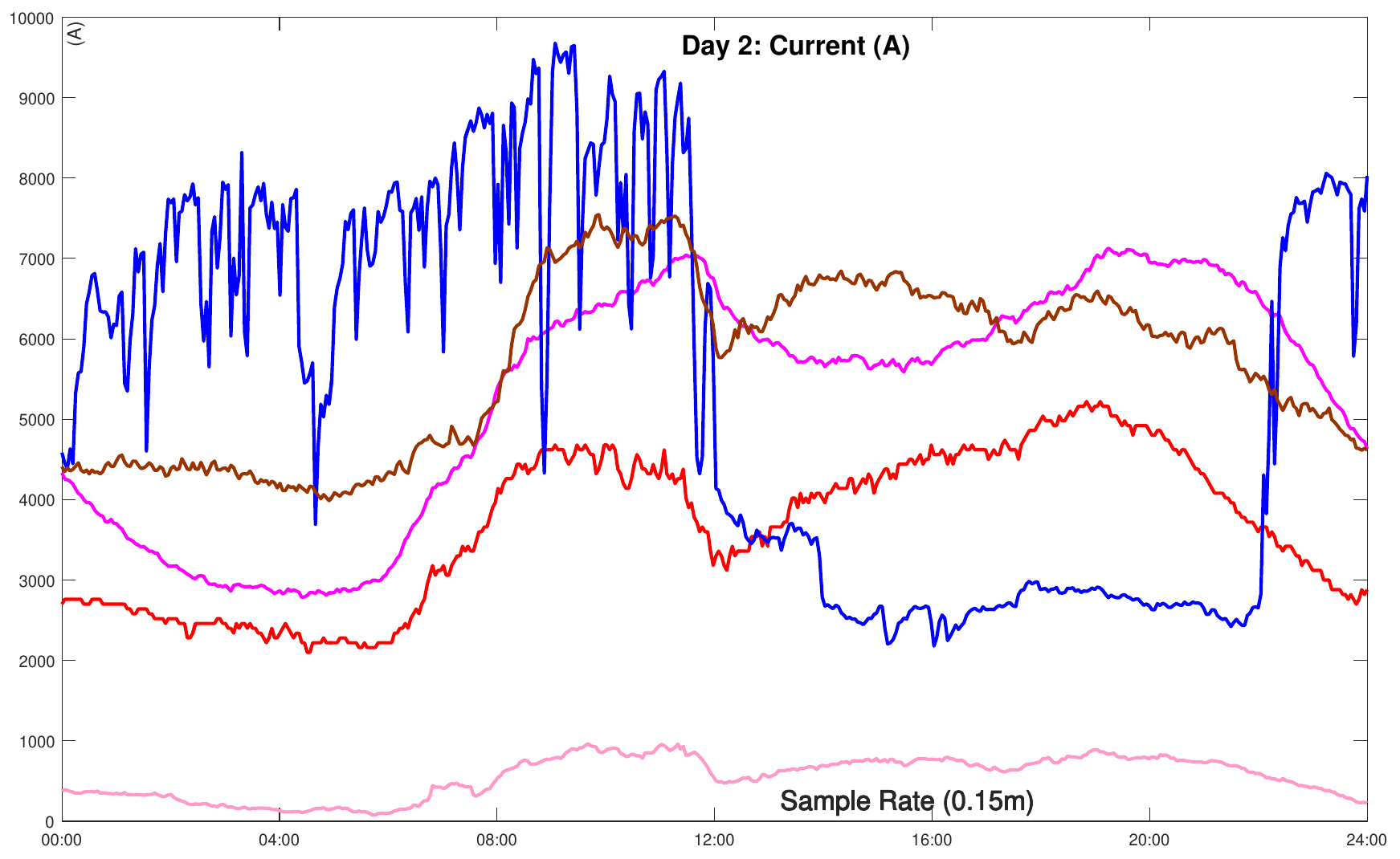}
}
\end{minipage}
\caption{Grid Network and Raw Data of Real Case}
\scriptsize {Note: For each substation, the 3-phase data are quite similar and only B-phase data are chosen.}
\normalsize{}
\label{fig:case5}
\end{figure*}

A power grid with 5 substations in China is studied in this section as shown in Fig.  \ref{Fig:Case5Gird}. For each substation, its three-phase voltage data $V$ and current data $I$ are recorded at a three-minute sampling-rate. We take a two-day time period as the dataset, depicted in Fig. \ref{fig:case5A1}, \ref{fig:case5A2}, \ref{fig:case5B1}, and \ref{fig:case5B2}.

\subsubsection{Ring Law and LES}  
\begin{figure*}[h]
\centering
\subfloat[Day 1: $t-\tau_{\text{T}_2}$]{\label{fig:case6A1}
\includegraphics[width=0.48\textwidth]{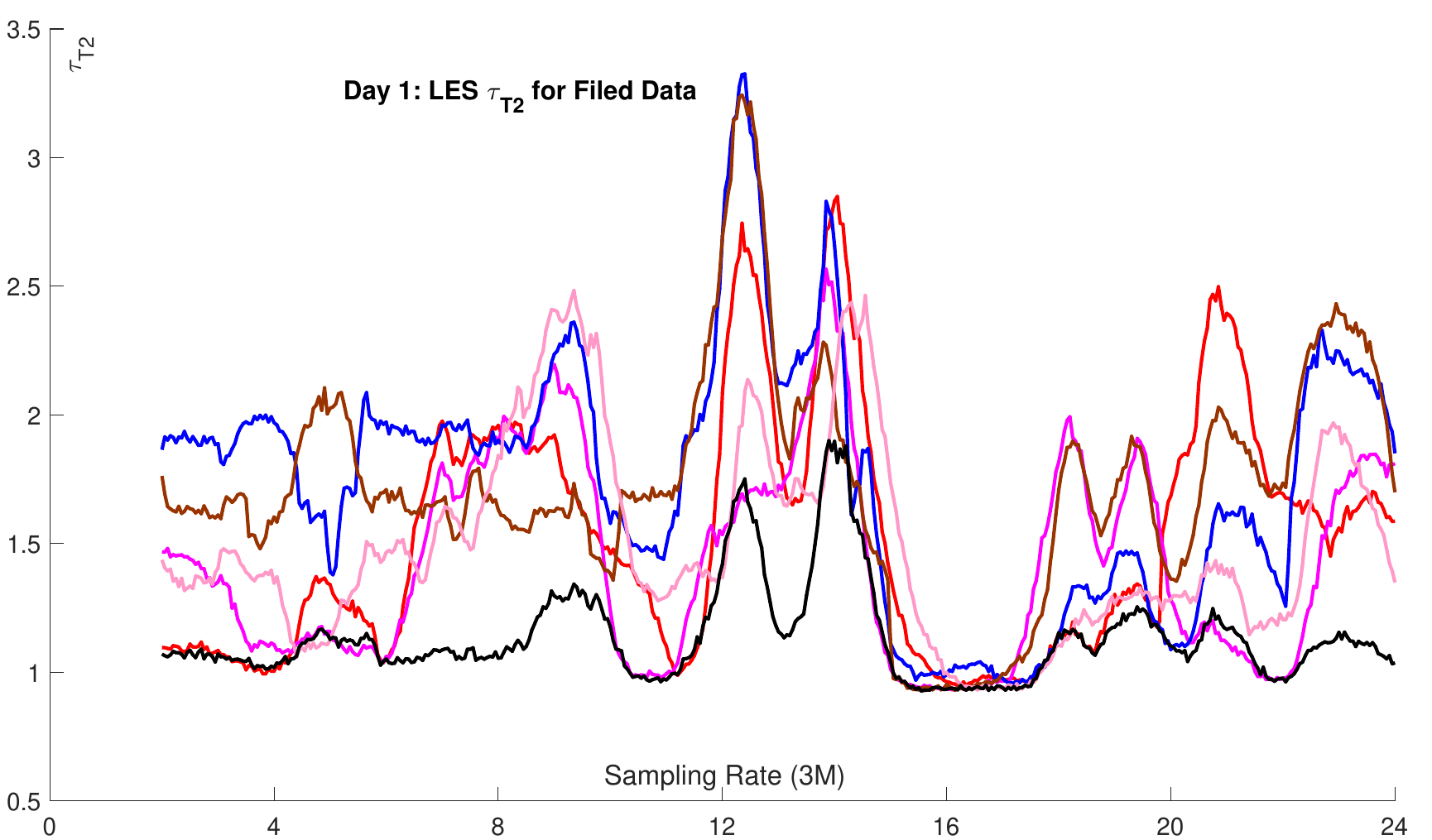}
}
\subfloat[Day 1: $t-\tau_{\text{LR}}$]{\label{fig:case6A2}
\includegraphics[width=0.48\textwidth]{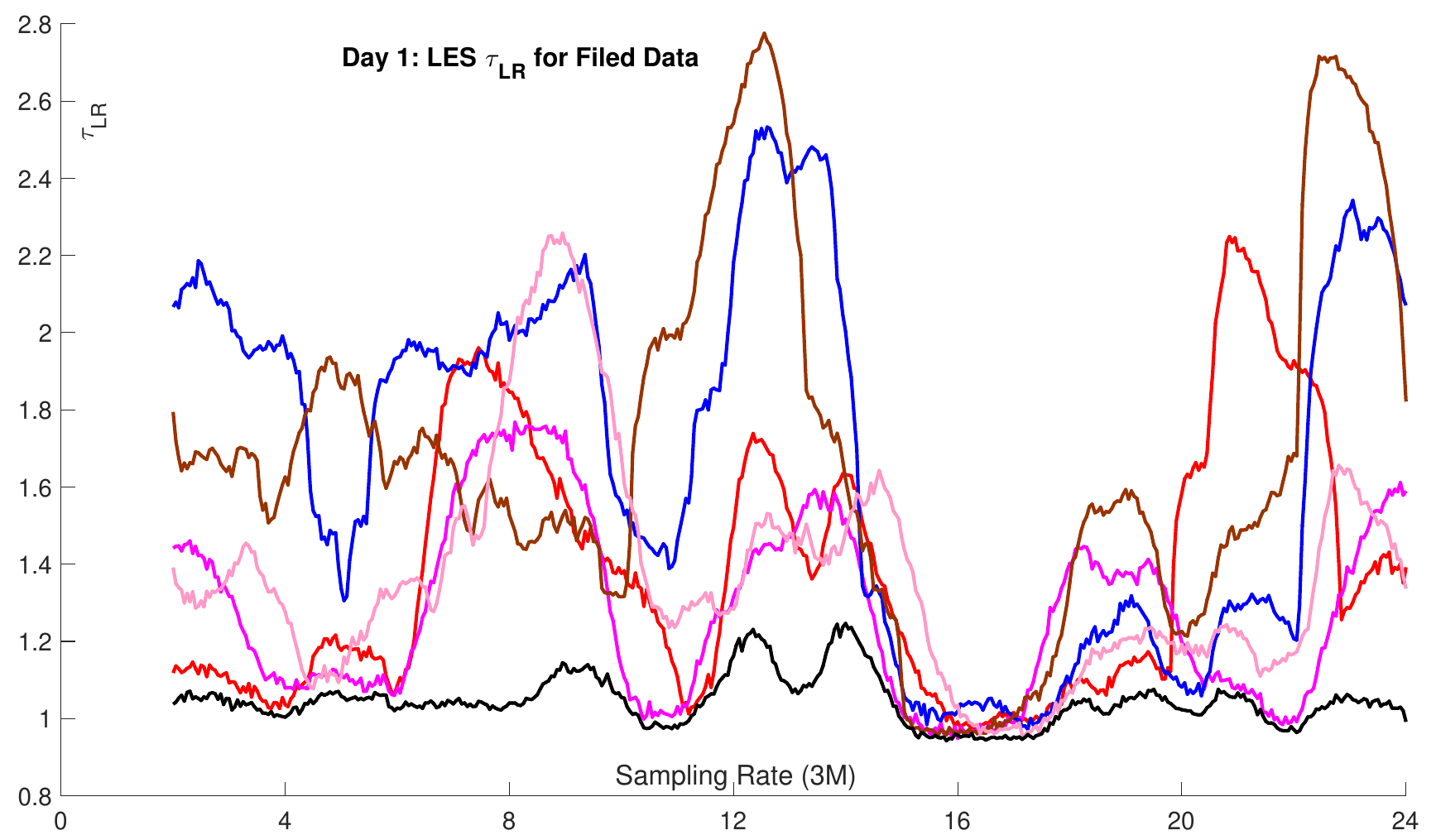}
}

\subfloat[Day 2: $t-\tau_{\text{T}_2}$]{\label{fig:case6B1}
\includegraphics[width=0.48\textwidth]{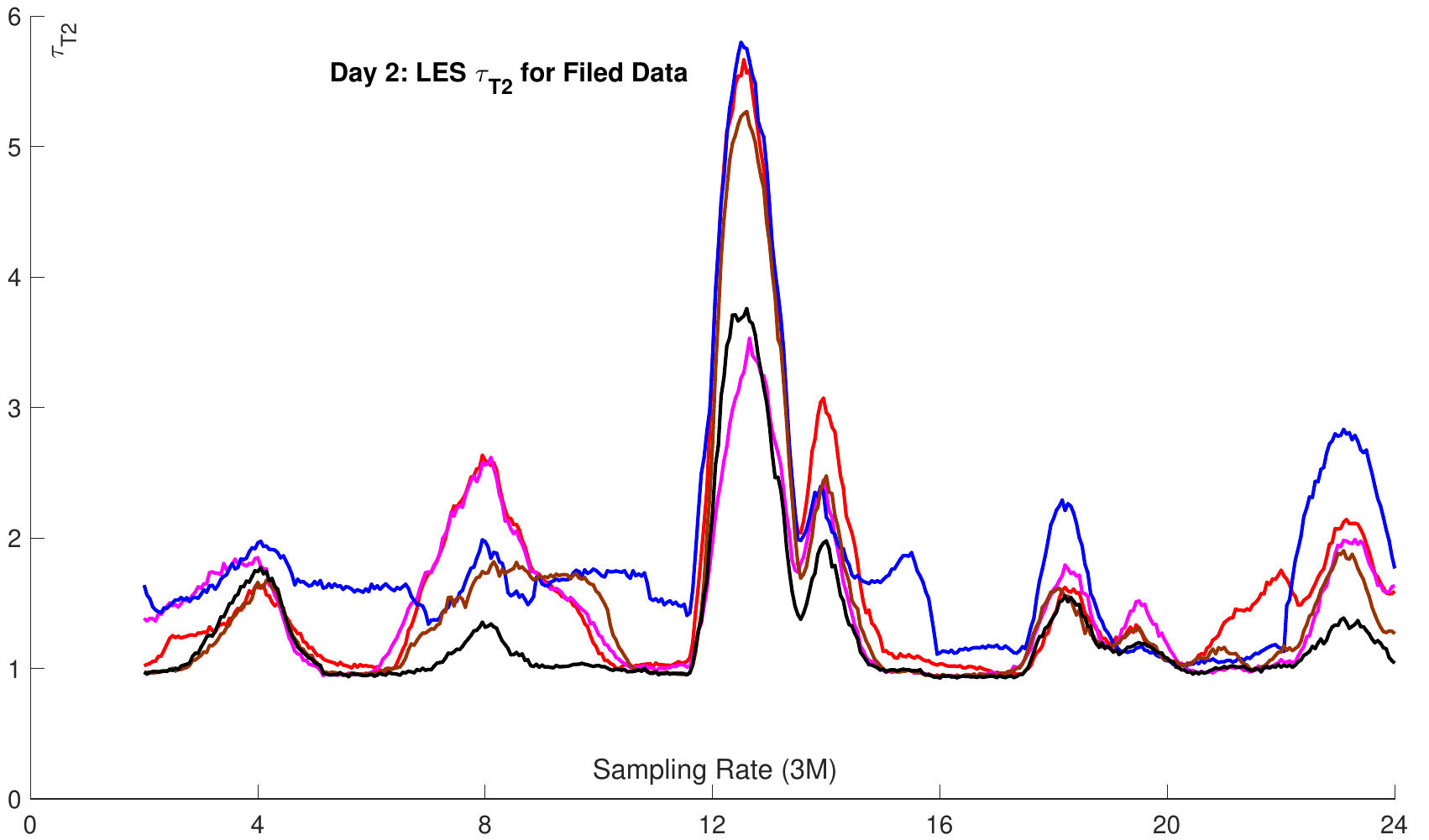}
}
\subfloat[Day 2: $t-\tau_{\text{LR}}$]{\label{fig:case6B2}
\includegraphics[width=0.48\textwidth]{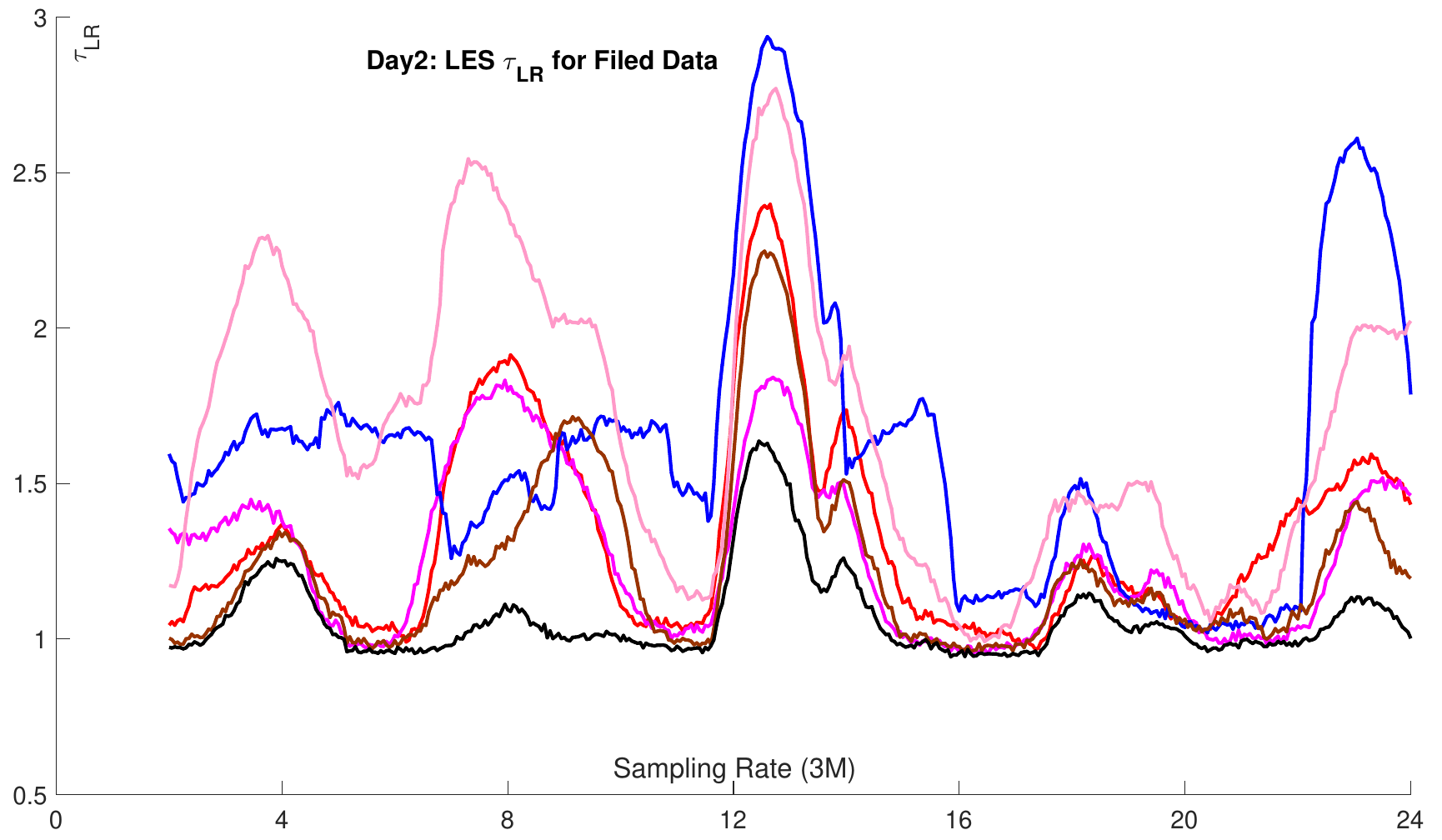}
}
\caption{Illusion of the LESs of field data.}
\label{fig:case6C}
\end{figure*}

If we choose $\mathbf{X}_0$ (the voltage data during 0 a.m. to 2 a.m, Fig. \ref{fig:case5A1}), the ring distribution is obtained according to our previous work \cite{he2015arch}, as shown in Fig.  \ref{fig:case5Ring}. Most eigenvalues are distributed between the inner circle and the outer circle. This implies that the real-world data do follow the Ring Law. With a similar process, and setting the test function as Chebyshev Polynomials  $\text{T}_2$: $\varphi_{\text{T}_2}(x)\!=\!2x^2\!-\!1$ and the Likelihood Ratio Function  $\text{LR}:\!{\varphi_\text{LR}(x)\!=\!x\!-\!\text{ln}(x)\!-\!1}$, respectively\footnote{Our previous work \cite{he2016les} discusses the effectiveness of  the choice of test functions. $\varphi_{\text{T}_2}(x)$ performs well on the calculation speed, and $\varphi_\text{LR}(x)$ performs well on the coefficient of variation in some cases.}, the LES $t\!-\!\tau$ curves are obtain in Fig. \ref{fig:case6A1}, \ref{fig:case6A2}, \ref{fig:case6B1}, and \ref{fig:case6B2}. The grid is relatively smooth during 0 a.m. to 8 a.m. and has dramatic changes at around 8:30 a.m., 11:30 a.m., etc. This observation agrees with our common sense. For  field data, the test function may influence the results in some complicated ways, although the LESs have a similar trend at most CPs.

\section{Conclusion}
\label{section: concl}
\normalsize{}

This paper extends our framework of using large random matrices to model a power grid. Under the RMT-based framework, a progress, aiming at the invisible units detection and estimation task, is made from the conventional techniques such as change point detection and least squares methods, and detailed algorithms are given.
Linear Eigenvalue Statistic (LES), whose value is robust against data errors and insusceptible to random noises, is employed as the statistic feature (or big data analytics). Due to the high-dimensional phenomenon---arising from considering a large number of nodes simultaneously, LES is very appealing  due to its Gaussian property implying that the expectation $\mathbb{E}(\tau)$ and the variance $\sigma^2(\tau)$ are sufficient for a complete statistic description.  %Employing LES, we exploit statistical properties of massive datasets in a high-dimensional vector space.
%Studying the statistical properties of LES, i.e., $\mathbb{E}(\tau)$ and $\sigma(\tau)$,
As a result, LES maps the massive datasets into a standard Gaussian random variable that is standard in classical statistics. The fundamental role of LES is highlighted here. As an example of how LES is used, we consider a standard (binary) hypothesis testing for invisible unit modeling. In particular, both the fraudulent behavior, e.g., false data injection, and anomalous power usage, e.g., unauthorized PV installation, are studied.

%The characteristic of our method is that we utilize some statistical properties of the observed data which are unique in high-dimensional space.
Based on mathematically rigorous RMT, time and space must be \textbf{tied together} through their ratio $c\!=\!T/N.$ What matters is the ratio $c,$ rather than $N$ and $T$. This observation is valid when $N$ and $T$ are large (for some statistical properties, size of tens is enough) and comparable in size, which is often true in practice.

Simulated data and real-world data are tested using our algorithms. It is found that the experimental values of LESs agree with the theoretical predictions. The proposed method does provide a data-driven approach to gain insight into the distribution network behavior and consumer profile, and a much more accurate estimation result is obtained.

%This paper we move forward to the detection and estimation of invisible units task, although there is still some distance left.

%
%The proposed RMT-based methodology  has numerous unique advantageous, and it is more suitable for complicated systems with easily accessible data.
%In the form of large random matrix, it handle massive data which are in high-dimension and within a wide time span all at once. In this way, highly reliable decisions are still attainable with some bad data, e.g., the unsynchronized data caused by erroneous time-tags or communication delays. Moreover, with the statistical processing such as test function setting, the proposed data-driven methodology has the potential to balance the perspectives of the speed, the sensitivity, and the reliability in practice.
%
%The stability evaluation and behavior analysis are two big topics along this direction. Besides, the statistical indicators are good medium for artificial intelligence and machine learning.

%

\bibliographystyle{IEEEtran}
\bibliography{helx1}


\normalsize{}
\end{document}